\documentclass[traditabstract]{aa}  
\usepackage{newtxmath}
\usepackage{newtxtext}
\usepackage{multirow}

\usepackage[T1]{fontenc}

\DeclareRobustCommand{\VAN}[3]{#2}
\let\VANthebibliography\thebibliography
\def\thebibliography{\DeclareRobustCommand{\VAN}[3]{##3}\VANthebibliography}

\usepackage{graphicx}	
\usepackage{amsmath}	
\usepackage[dvipsnames,usenames]{color} 
\usepackage{longtable}
\usepackage{float} 
\usepackage{natbib}
\usepackage{adjustbox}
\usepackage[normalem]{ulem}
\usepackage{placeins}

\newcommand{\rmd}{{\rm d}}

\newcommand{\ie}{{i.e.~}}
\newcommand{\cf}{{cf.~}}
\newcommand{\eg}{{e.g.~}}

\begin{document} 

\title{Buoyancy glitches in pulsating stars revisited}

\author{Margarida S. Cunha\inst{1,2} \and Yuri C. Damasceno\inst{1,3} \and  Juliana Amaral\inst{1,3} \and Anselmo Falorca\inst{1,3} \and  J\o rgen Christensen-Dalsgaard\inst{4} \and Pedro P. Avelino\inst{1,2,3}}
 \institute{Instituto de Astrof\'{\i}sica e Ci\^{e}ncias do Espa\c{c}o, Universidade do Porto, CAUP, Rua das Estrelas, PT4150-762 Porto, Portugal\\ \email{mcunha@astro.up.pt} \and Université Côte d'Azur, Observatoire de la Côte d'Azur, CNRS, Laboratoire Lagrange, Bd de l'Observatoire, CS 34229, 06304 Nice cedex 4, France \and Departamento de Física e Astronomia, Faculdade de Ciências, Universidade do Porto, Rua do Campo Alegre 687, PT4169-007 Porto, Portugal \and Stellar Astrophysics Centre, Department of Physics and Astronomy, Aarhus University, Ny Munkegade 120, DK-8000 Aarhus C, Denmark}

\date{Accepted XXX. Received YYY; in original form ZZZ}



\abstract
{Sharp structural variations induce specific signatures on stellar pulsations that can be studied to infer localised information on the stratification of the star. This information is key to improve our understanding of the physical processes that lead to the structural variations and how to model them. Here we revisit and extend the analysis of the signature of different types of buoyancy glitches in gravity-mode and mixed-mode pulsators presented in earlier works, {including glitches with step-like, Gaussian-like, and Dirac-$\delta$-like shapes.} In particular, we provide analytical expressions for the perturbations to the periods and show that these can be reliably used in place of the expressions provided for the period spacings, with the advantage that the use of the new expressions does not require modes with consecutive radial orders to be observed. Based on a comparison with two limit cases and on simulated data, we further tested the accuracy of the expression for the  Gaussian-like glitch signature whose derivation in an earlier work involved a significant approximation. We find that the least reliable glitch parameter inferred from fitting that expression is the amplitude, which can be up to a  factor of two larger than the true amplitude, reaching this limit when the glitch is small. We further discuss the impact on the glitch signature of considering a glitch in the inner and outer half of the g-mode cavity, emphasising the break of symmetry that takes place in the case of mixed-mode pulsators.}


\maketitle


\section{Introduction}
\label{sec:intro}

Space-born experiments, such as the Convection, Rotation and planetary Transits \citep[CoRoT][]{baglin06}, Kepler \citep{Kepler10,gillilandetal10}, and the Transiting Exoplanet Survey Satellite \citep[TESS][]{TESS15} have fuelled the field of stellar structure and evolution by providing exquisite data sets on stellar pulsations that are explored via asteroseismology \citep{aerts10,cunhaetal07,cunha18}. One of the promising methods employed to infer details of the internal structure and dynamics of stars is based on the study of glitch signatures, that is, the signatures left  on the frequencies or periods of the oscillations by regions of sharp structural variations. Here `sharp' means that the characteristic scale of the structural variation is comparable to or smaller than the local wavelength of the waves to be studied \citep[see][for a review]{cunha20}. 

The potential of studying signatures of acoustic glitches associated with the base of the convective envelope and the helium ionisation zone has been recognised since the early days of helioseismology \citep[e.g.][]{hill86,thompson1988,GoughandThompson1988,vorontsov1988,gough90}. In this context, numerous theoretical and observational studies have been published focussing on the Sun \citep[e.g.][]{basu94,monteiroetal94,roxburgh94,JCD95,basu97,monteiroetal00,monteiro05,houdek07,roxburgh09,jcd11} and on other solar-like pulsators, including main-sequence \citep[e.g.][]{roxburgh01,ballot04,basu04,lebreton12,mazumdar14,verma14,verma17,verma19,farnir19,deal23} and red-giant stars \citep{miglio10,broomhall14,corsaro15,dreau20,saunders23}.

The potential of using acoustic modes to extract information on sharp structural variations located in stellar cores has also been discussed in the literature \citep[e.g.][]{audard94,roxburgh99,roxburgh01,mazumdar06,cunha07,cunha11,lindsay23}, although the realisation of this potential proved more challenging, particularly for deep glitches such as those associated with small convective cores.  A more promising approach to detect the seismic impact of these core structural variations
is to look for glitch signatures in the periods or frequencies of gravity (g) or mixed modes, when these are available. 
The seismic signatures of such buoyancy glitches and the mode trapping they induce have also received the attention of numerous works, including works
focussing on white dwarfs \citep[e.g.][]{winget81,winget91,brassard92,benvenuto02,corsico02}, intermediate and massive main-sequence stars \citep[e.g.][]{miglio08,degroote10,kurtz14,vanreeth2015,mombarg22}, subdwarf O and B (sdO, sdB) stars \citep[e.g.][]{charpinet00,charpinet02,rodriguez10,ostensen14,baran17,Ghasemi17,Uzundag17},  and red-giant stars \citep[e.g.][]{mosser15,cunha15,cunha19,jiang22,vrard22}. Recently, the theoretical problem has been addressed in a more general angle by considering the eigenvalue condition built up from a series of resonant cavities \citep{pinçon22}.

Different methodologies can be considered for extracting information contained in glitch signatures. A common approach is to produce a set of stellar models by varying particular aspects of the stellar physics in an attempt to reproduce the glitch signature. An alternative approach of greater relevance to the present work is to develop an analytical description of the glitch signature that enables the extraction of the glitch properties (\eg location and amplitude) directly from fitting the data, without having to recur to specific stellar models. This `model-independent' approach was considered in the context of main-sequence g-mode pulsators by \cite{miglio08} under the assumption that the impact of the glitch on the pulsation periods can be treated as a small perturbation (\ie using a variational approach), which is often not the case. When the glitch impact on the pulsation periods is not small, perturbations to the pulsation periods may be derived  instead by matching the wave solutions on each side of the glitch. This approach was used by
\cite{cunha15,cunha19} to derive analytical expressions for the period spacings of g-mode and mixed-mode pulsators in the presence of buoyancy glitches with different shapes, namely, Dirac-$\delta$-like, step-like, and Gaussian-like glitches.  Following the same approach, \cite{hatta23} has recently considered, in addition, the impact on the period spacings of g-mode pulsators of a buoyancy glitch with a ramp-like shape.

{Period spacings can only be computed when modes of consecutive radial order are observed. However, often not all radial orders in a sequence are detected \cite[e.g.][]{ostensen14}, while enough  information is still available to estimate the asymptotic period spacing. In these cases, one would expect that fitting the observed periods directly would be a good alternative, regardless of them being consecutive or not. This also has the advantage of avoiding approximations introduced in the analytical analysis presented in earlier works when relating the perturbation in the periods to the perturbations in the period spacings (cf. equation 34 in \cite{jcd12}).
With this in mind, in the present work we revisit the step-like and Gaussian-like cases to provide explicit analytical expressions for the period perturbations that can be used when not all modes in a sequence of radial orders are observed, but an estimate of the asymptotic period spacing is still available. }

Matching the wave solutions on each side of the glitch is relatively straightforward when the glitch can be assumed to be infinitely thin, for example when it is adequately modelled by a Dirac-$\delta$ function or a step function.  However, the derivation becomes less accurate when the glitch has a non-negligible width. Matching across the glitch in this case requires an analytical representation of the eigenfunction inside the glitch, which, in turn, cannot be derived asymptotically because the scale of variation of the background there becomes comparable to the scale of the wave. Therefore, in the present work, we also explain how we performed additional tests to the expression provided earlier for the Gaussian-like glitch \citep{cunha19} to quantify the impact of the assumptions that underlined its derivation.



We start by providing a general analytical expression for the period perturbations of pure g modes in Section~\ref{sec:periods}. 
We then consider the case of a step-like glitch (Section~\ref{sec:step}), testing the perturbation to the periods against model data and verifying that the perturbation to the periods in the limit of a small amplitude glitch is identical to that derived with recourse to the variational principle, as expected. Next, we address the case of the Gaussian-like glitch, discussing the limits of the proposed periods' perturbation when the glitch approaches the Dirac $\delta$ case and the case of a small glitch, and setting the accuracy limits of the expression based on artificial data produced to this effect. In Section~\ref{sec:additional} we discuss other cases, including the case of a buoyancy glitch described by a ramp function and the case of buoyancy glitches in stars exhibiting mixed modes. Finally, in Section~\ref{sec:conclusions} we draw the main conclusions from our work.

\section{Perturbation to the periods}
\label{sec:periods}
Internal gravity waves propagate in regions that are convectively stable with a phase speed that depends directly on the buoyancy frequency $N$. Asymptotically, at low degree ($l$) and high radial orders ($n$) the periods of gravity eigenmodes (hereafter { g modes}) are expected to be equally spaced  with a period spacing given by \citep{tassoul80}
\begin{equation}
\Delta P_{\rm as}= \frac{2\pi^2}{\omega_{\rm g}},
\label{psasymp}
\end{equation}
where,
$\omega_{\rm g}$ is the buoyancy extent of the g-mode cavity given by
\begin{equation}
\omega_{\rm g}=\int_{r_1}^{r_2}\frac{LN}{r}\rmd r,
\end{equation}
$L=\sqrt{l(l+1)}$, $r$ is the distance from the stellar centre, and $r_1$ and $r_2$ are the inner and outer turning
points, respectively, of the g mode propagation cavity.

When a glitch is located inside the g-mode cavity, the phase of the wave is perturbed, which in turn perturbs the eigenvalue condition and shifts the periods from the constant spacing predicted asymptotically. 
Under the Cowling approximation and using the variable $\Psi=~(r^3/g\rho \tilde f) ^{1/2}\delta p$, where $\delta p$ is the Lagrangian pressure perturbation, $\rho$ is the density, $g$ is the gravitational acceleration, and $\tilde f$ is a function of frequency and of the equilibrium structure [the f-mode discriminant defined by equation 35 of \cite{gough07}], the wave equation resulting from the linear, adiabatic pulsation equations, for the case of a spherically symmetric equilibrium can be written in the standard form as
\begin{eqnarray}
{\Psi^{\prime\prime}}+K^2\Psi=0.
\label{waveeqap}
\end{eqnarray}
Here a prime represents  a differentiation with respect to $r$ and, given our interest in pure g modes, we shall approximate the radial wavenumber $K$ by
 \begin{equation}
  K^2\approx -\frac{L^2}{r^2}\left(1-\frac{N^2}{\omega^2}\right),
\label{k2}   
 \end{equation}
with $\omega$ being the angular frequency of the mode.

Starting from equation (\ref{waveeqap}) and following the asymptotic analysis by \cite{gough93}, \cite{cunha19} showed that the perturbed eigenvalue condition in the presence of a buoyancy glitch takes the form,
\begin{equation}
\sin\left(\int_{r_1}^{r_2}K\rmd  r+\frac{\pi}{2}+\Phi\right) = 0,
\label{eigencondition}
\end{equation}
 where  $\Phi$ is the frequency-dependent phase perturbation induced by the glitch.

From equation (\ref{eigencondition}), it follows that,
\begin{equation}
    \int_{r_1}^{r_2}K(\omega)\rmd  r=\pi\left(n-\frac{1}{2}\right)-\Phi(\omega),
\end{equation}
where $n$ is an integer.

To derive the perturbation to the oscillation periods, with regards to a reference model with a buoyancy frequency $N_0$, we start by noticing that in the reference model the eigenvalue condition implies that, 
\begin{equation}
\int_{r_1}^{r_2}K_0(\omega_0)\rmd  r = \pi\left(n-\frac{1}{2}\right),
\label{eigencondition0}
\end{equation}
where $\omega_0$ is the unperturbed eigenfrequency, and thus,  
\begin{equation}
  \int_{r_1}^{r_2}K(\omega)\rmd  r= \int_{r_1}^{r_2}K_0(\omega_0)\rmd  r -\Phi\left(\omega\right). 
  \label{int}
\end{equation}
Except near the turning points, $K\approx LN/r\omega$ and $K_0\approx LN_0/r\omega_0$.  Therefore one can write $\int_{r_1}^{r_2}K(\omega)\rmd  r=(\int_{r_1}^{r_2}LN/r\omega \rmd  r+\alpha$) and $\int_{r_1}^{r_2}K_0(\omega_0)\rmd  r=(\int_{r_1}^{r_2}LN_0/r\omega_0 \rmd  r+\alpha$) where $\alpha=\delta+\tilde\delta$ is a phase that incorporates the effect of the approximation of the radial wavenumber near both turning points ($\delta$ and $\tilde\delta$ being the contributions from the outer and inner turning points, respectively). Notice that we have taken the turning points and $\alpha$ to be the same in the perturbed and unperturbed models. This is justified by the fact that the buoyancy frequency is unperturbed near the turning points (or else, the perturbation would not be a glitch, because there the wavenumber tends to zero and the glitch condition would not be satisfied). Thus, the turning points and $\alpha$ in the perturbed and unperturbed cases may differ only due to the change in the eigenfrequency, which in all cases considered is small (in relative terms), even when the local perturbations to $N$ inside the g-mode cavity are not. 

Written in terms of the periods $P=2\pi/\omega$, it then follows from equation (\ref{int}) that the perturbation to the periods with regards to the reference model is,
\begin{equation}
 P=P_0-P_0\left(\int_{r_1}^{r_2}\delta N \frac{\rmd r}{r}\right)\left(\int_{r_1}^{r_2}N\frac{\rmd r}{r}\right)^{-1}-\left(\int_{r_1}^{r_2}\frac{LN}{2\pi}\frac{\rmd r}{r}\right)^{-1}\Phi,
 \label{period}
\end{equation}
where we introduced the perturbation to the buoyancy frequency with respect to a reference model, defined by $\delta N=N-N_0$. Notice that this perturbation depends on the choice of reference model, which can take different forms. An example of such a reference model is given in Appendix~\ref{apA} for the case of a step-like glitch.

From equation (\ref{period}) one can identify two contributions to the perturbed period $\delta P=P-P_0$: a smooth contribution expressed by the second term on the right hand side (rhs) of the equation, related to the change in the total integral of the buoyancy frequency within the g-mode cavity, and the contribution from the glitch-induced phase $\Phi$.  

Equation (\ref{period}) can also be written in terms of the asymptotic period spacing, $\Delta P_{{\rm as}}$, namely,
\begin{equation}
    P=P_0-P_0\frac{\Delta P_{\rm as}}{\pi}\left(\int_{r_1}^{r_2}\frac{L\delta N}{2\pi} \frac{\rmd r}{r}\right)-\frac{\Delta P_{\rm as}}{\pi}\Phi\equiv P_{\rm s}-\frac{\Delta P_{\rm as}}{\pi}\Phi,
 \label{period_as}
\end{equation}
where we have introduced $P_{\rm s}$ (on the rightmost side of the expression)  representing the unperturbed periods ({\ie} without the glitch effect) of a hypothetical model in which the integral of $N/r$ within the g-mode cavity is the same as in the glitch model. {Asymptotically, we can write $P_{\rm s} = P_{\rm s,min}+k\Delta P_{\rm as}$, for a series of natural numbers $k$, where $P_{\rm s,min}$ is the first of a series of equally spaced unperturbed periods. Rewriting equation (\ref{period_as}), we then have,
\begin{equation}
    P= P_{\rm s,min}+\Delta P_{\rm as}\left(k-\frac{\Phi}{\pi}\right).
 \label{period_as2}
\end{equation}
Therefore, $\Phi/\pi$ is directly related to the deviation of the periods from the exact asymptotic spacing and its impact is seen as a modulation in a period échelle diagram.  }

As the reference model is not a priori univocally defined, we shall consider the period perturbations with respect to $P_{\rm s}$ ($\delta P_{\rm s}=P-P_{\rm s}$), rather than the $\delta P$ defined above, unless otherwise stated.
It is worth noting that when the asymptotic period spacing is the same in the glitch model and in the chosen reference model, $P_{\rm s}$ and $P_0$ are also the same. 

Finally, as the asymptotic period spacing depends on the mode degree, to fit modes of different degree simultaneously it is useful to rewrite the expression above in terms of the $l$-independent reduced period spacing $\Delta\Pi=L\Delta P_{\rm as}$. Defining the reduced period $\Pi=L P$ one then finds that
\begin{equation}
    \Pi=\Pi_0-\Pi_0\frac{\Delta \Pi_{\rm as}}{\pi}\left(\int_{r_1}^{r_2}\frac{\delta N}{2\pi} \frac{\rmd r}{r}\right)-\frac{\Delta\Pi_{\rm as}}{\pi}\Phi\equiv \Pi_{\rm s}-\frac{\Delta\Pi_{\rm as}}{\pi}\Phi,
 \label{period_as_red}
\end{equation}
with $\Pi_{\rm s}=LP_{\rm s}$.

In what follows we shall consider the explicit form taken by the glitch phase, $\Phi$, for different representations of the glitch.  To that end, we define the buoyancy radius
\begin{equation}
    \tilde\omega_{\rm g}^r=\int_{r_1}^{r}\frac{LN}{r}\rmd r
\end{equation}
and the buoyancy depth
\begin{equation}
    \omega_{\rm g}^r=\int_{r}^{{r}_2}\frac{LN}{r}\rmd r,
\end{equation}
and note that both these quantities are defined in terms of the buoyancy frequency $N$ in the star or model under consideration and not in terms of the buoyancy frequency $N_0$ of the reference model. This is of particular relevance in the case of finite-width glitches, such as a glitch represented by a Gaussian function.
The location of glitches in the inner half of the g-mode cavity (\ie for which $\tilde\omega_{\rm g}^r/\omega_{\rm g} < 0.5$) shall be expressed in terms of the buoyancy radius while the location of glitches in the outer half of the g-mode cavity shall be expressed in terms of the buoyancy depth.  

\section{Step-like glitch}
\label{sec:step}
\cite{cunha19} considered the case of a decreasing step-like glitch in the inner half of the g-mode cavity, defined by 
\begin{equation}
N=\left\{
\begin{array}{lll}
 N_{\rm in}  &   {\rm for} & r < r^\star\\
 N_{\rm out}  & {\rm for} & r > r^\star \\
\end{array}
\right. ,
\label{brunt}
\end{equation}
with $N_{\rm in}>N_{\rm out}$, thus with $N$ varying by the positive amount $\Delta N= \left. N_{\rm in}\right|_{ r\rightarrow r^\star_-}-\left. N_{\rm out}\right|_ {r\rightarrow r^\star_+}$ at $r=r^\star$ (not to be confused with the perturbation with respect to the reference model, $\delta N$).
{According to the authors (and keeping their notation), the glitch-induced phase for this case is given by,
\begin{equation}
       \Phi_{\rm st}={\rm arccot}\left[-\frac{2}{A_{\rm st}\sin\left(2\tilde\beta_2\right)}-\cot\tilde\beta_2\right],
        \label{phi_minus}
\end{equation}
where the subscript ``st'' is used to indicate a step-like glitch. Here,
$\tilde\beta_2=\int_{r_1}^{r^\star}K\rmd  r +\pi/4 \approx \frac{\tilde\omega_{\rm g}^\star}{2\pi} P+\pi/4+\tilde\delta$, and the superscript $^\star$ indicates that the corresponding quantity is evaluated at $r=r^\star$.} The glitch is thus characterised by the relative step amplitude
$A_{\rm st}=[N_{\rm in}/N_{\rm out}]_{r^\star}-1$, and the buoyancy radius $\tilde\omega_{\rm g}^\star$ at the glitch location, with the expression including an additional fudge parameter $\tilde\delta$ accounting for the approximation of the integral made near the inner turning point. While \cite{cunha19} considered a model with a decreasing step-like glitch, the authors did not make any assumption about the sign of $\Delta N$ when deriving equation~(\ref{phi_minus}). Therefore, this equation is also valid for an increasing step-like glitch for which $\Delta N<0$. 

It should be noted that the expression for $\Phi_{\rm st}$ is degenerate with respect to the transformation $A_{\rm st}\rightarrow \hat A_{\rm st}$ and $\tilde\delta \rightarrow \tilde\delta-\pi/2$, where $\hat A_{\rm st}=[N_{\rm out}/N_{\rm in}]_{r^\star}-1$. Given that $A_{\rm st}$ and $\hat A_{\rm st}$ have opposite signs and are not identical in absolute value, care must be exerted in the interpretation of the retrieved amplitude. In the case of a decreasing step-like glitch with a fixed jump in $N$ at the discontinuity, the positive amplitude retrieved from the fitting will be larger than the absolute value of the negative solution. This is because the former should be interpreted as $A_{\rm st}$ and the latter should be interpreted as $\hat A_{\rm st}$ (the actual jump being the same, but expressed differently). The situation is reversed for an increasing step-like glitch. This difficulty can be avoided by setting a prior to the amplitude when fitting the data. By imposing that the amplitude  must be positive the retrieved value is always interpreted as $A_{\rm st}^{\pm}=[N_{\rm +}/N_{\rm -}]_{r^\star}-1$, where the subscripts $+$ and $-$ indicate the largest and the smallest of the two values of $N$ at the discontinuity.

 Finally, while the expression above was derived for a glitch located in the inner half of the g-mode cavity, by symmetry of the mathematical problem considered, it is clear that the case of a decreasing step-like glitch in the inner cavity corresponds to the case of an increasing step-like glitch in the outer cavity, and the case of an increasing step-like glitch in the inner cavity corresponds to the case of a decreasing step-like glitch in the outer cavity. Thus, to infer the properties of a glitch in the outer g-mode cavity one can also use expression (\ref{phi_minus}), making the correspondence $\tilde\beta_2 \rightarrow\beta_2$, where $\beta_2=\int_{r^\star}^{r_2}K\rmd  r +\pi/4\approx \frac{\omega_{\rm g}^\star}{2\pi} P+\pi/4+\delta$. 
 Notice, however, that the phases $\tilde\delta$ and $\delta$ depend on the details of $K$ near the inner and outer turning points, respectively, and are not necessarily the same. Hence, despite the symmetry of the expression, the signatures of glitches located at equal values of $\omega_{\rm g}^\star$ and $\tilde\omega_{\rm g}^\star$ may differ.

\subsection{Test on model data \label{sec:step_tests}}

{Massive stars have convective cores that retreat as they evolve in the main sequence. The retreating core leaves behind a gradient in the hydrogen abundance that changes abruptly at the radius where the mixing region was once at its maximum. This abrupt change in the gradient of hydrogen abundance leads to a step-like discontinuity in the buoyancy frequency that is located inside the g-mode cavity, hence ideal to test our formulation \cite[see, \eg figure 1, left panels, of][for the chemical and buoyancy profiles in a massive star]{cunha19}.} To test the expression for the periods in the presence of a step-like glitch we, thus, fit equation (\ref{period_as}) expressed in terms of the unperturbed periods $P_{\rm s}$, to the periods computed for a model of a main-sequence, 6~M$_\odot$ star. {In order to compare with the results obtained by \cite{cunha19}, we use the model and adiabatic pulsation periods computed in their work using the Aarhus STellar Evolution Code \citep[ASTEC,][]{jcd08a} and the Aarhus adiabatic oscillation package  \citep[{ADIPLS,}][]{jcd08b}, respectively. }

The model data were fitted through a Markov Chain Monte Carlo method (MCMC) using the emcee python package, where the likelihood was assumed to be Gaussian, as stated in expression (\ref{eq:likelihood}), where $\sigma$ represents the jitter parameter.
\begin{equation}
    \mathcal{L} = \left(\frac{1}{\sqrt{2\pi\sigma^2}}\right)^N \exp\left[-\frac{1}{2}\sum_i^N\left(\frac{P_i-P_{{\rm\small ADIPLS}, i}}{\sigma}\right)^2\right].
    \label{eq:likelihood}
\end{equation}

{As in previous works, we do not perturb the pulsation periods prior to performing the fit. Instead, we use the model periods as given and take $\sigma$ as a free parameter. Hence, the adequacy of the fit is reflected in the ratio of the inferred value of $\sigma$, to a characteristic value of the periods. }

Prior to the MCMC, the likelihood was maximised using a Nelder-Mead minimiser method on $-\log(\mathcal{L})$. The resulting parameter space position was used to set the MCMC walkers' initial position, adding to each a small (around 0.01\% of each parameter's value) random dislocation, allowing a better exploration of the parameter space near this region.

The construction of the fitted periods through equation (\ref{period_as}) involves the resolution of a transcendental equation in the periods. This was achieved using the bisection method, iteratively for each period, with a margin of $10^{-8}$. The codomain of the arc-cotangent in equation (\ref{phi_minus})  must be carefully selected in order to avoid discontinuities. For step function glitches the argument passes from $+\infty$ to $-\infty$, so the codomain was chosen to be $[-\pi/2, \pi/2]$.  With the discontinuities avoided all together, the determination of each period comes from finding the root of a monotonous function, which is doable by setting the search region as $[P_{i-1},P_{i-1}+2\Delta P_{as}]$. {The free parameters determined  from the fit are  $P_{\rm s,min}$, corresponding to the first of the periods $P_{\rm s}$, }the asymptotic period spacing $\Delta P_{\rm as}$, the jitter $\sigma$ and the glitch parameters, namely, the amplitude $A_{\rm st}$, position $\tilde\omega_g^\star$ and phase $\tilde\delta$.


We generated Markov Chains with 50 walkers and 9000 iterations, with a burn-in of 1000 iterations. From these we extracted the maximum probability parameters as well as the probability distribution median parameters and their respective 68\% confidence interval. {The priors on the parameters were uniform within the intervals provided in Table~\ref{tab:prior} and the respective marginalised distributions for the parameters are shown in Fig.~\ref{fig:corner_step}.}

Figure~\ref{fig:echelle_MS_RGB}, left panel, shows the periods of the {\small ASTEC} main-sequence model in the form of an échelle diagram, where the model periods (red symbols) are plotted against the {residuals defined as follows: 
\begin{eqnarray}
    {\rm res}=\left(P-P_{\rm s,min}\right)\mod\Delta P_{\rm as};&&\nonumber\\
    {\rm if} \hspace{0.3cm} {\rm res} > \Delta P_{\rm as}/2 \hspace{0.3cm}{\rm then} \hspace{0.3cm}{\rm res}={\rm res}-\Delta P_{\rm as} . &&
\end{eqnarray}
 The period perturbations  $\delta P_{\rm s} = -\frac{\Delta P_{\rm as}}{\pi}\Phi_{\rm st}$  derived from fitting equation~(\ref{period_as}) to the model periods, with the glitch phase defined by (\ref{phi_minus}), are also shown (black line).  The parameters inferred from the fit are given in Table~\ref{tab:parameters_step}. 
Comparison with the results found by \cite{cunha19} from fitting the periods spacings shows agreement within 1-$\sigma$ on the inferred glitch amplitude, glitch position, and period spacings. The phases also show a reasonable agreement ($\sim$~1.5-$\sigma$). 

\begin{table*}
	\caption{{Parameters derived from the fit of the analytical expression for the step-like glitch [equations~(\ref{period_as}) and (\ref{phi_minus})]  to the periods derived from {\small ADIPLS} for the main-sequence model.  The values shown correspond to the median of the distributions and the $68\%$ confidence intervals. For comparison, the values of the glitch parameters and asymptotic period spacing inferred by fitting the period spacings, as well as the glitch parameters estimated directly from the {\small ASTEC} buoyancy frequency profile are also shown.}}
	\label{tab:parameters_step}
 \centering
\label{parameters_step}
\begin{minipage}{0.7\textwidth}
	\resizebox{\linewidth}{!}{%
	\begin{tabular}{|c|c|c|c|c|c|c|}
			\hline
			&$P_{\rm s,min}$&$\Delta P_{\rm as}$ (s) & $A_{\rm st}$  &$\tilde\omega_{\rm g}^\star$ ($10^{-6}$ rad/s) & $\tilde\delta$ & reference\\
			\hline
			& & & & & &\\
     Periods' fit & $50315^{+176}_{-174}$ &{$8485^{ +12}_{-12}$}&{$4.71^{ +0.90}_{-0.72}$} &{$348.0^{ +2.8}_{-2.8}$}  &{$0.713^{ +0.076}_{-0.073}$} & this work\\
			& & & & & &\\
			Period Spacings' fit&&{$8472^{ +50}_{-50}$}&{$4.74^{ +0.44}_{-0.39}$} &{$351.61^{ +0.73}_{-0.72}$}  &{$0.602^{ +0.019}_{-0.019}$} & \cite{cunha19}\\
			& & & & & &\\
			\hline
			Estimated &--&--&{$5.3$}  & {349}&-- &\cite{cunha19}\\
			\hline
		\end{tabular}
  }
\end{minipage}
\end{table*}


\begin{figure}
	\includegraphics[width=\columnwidth]{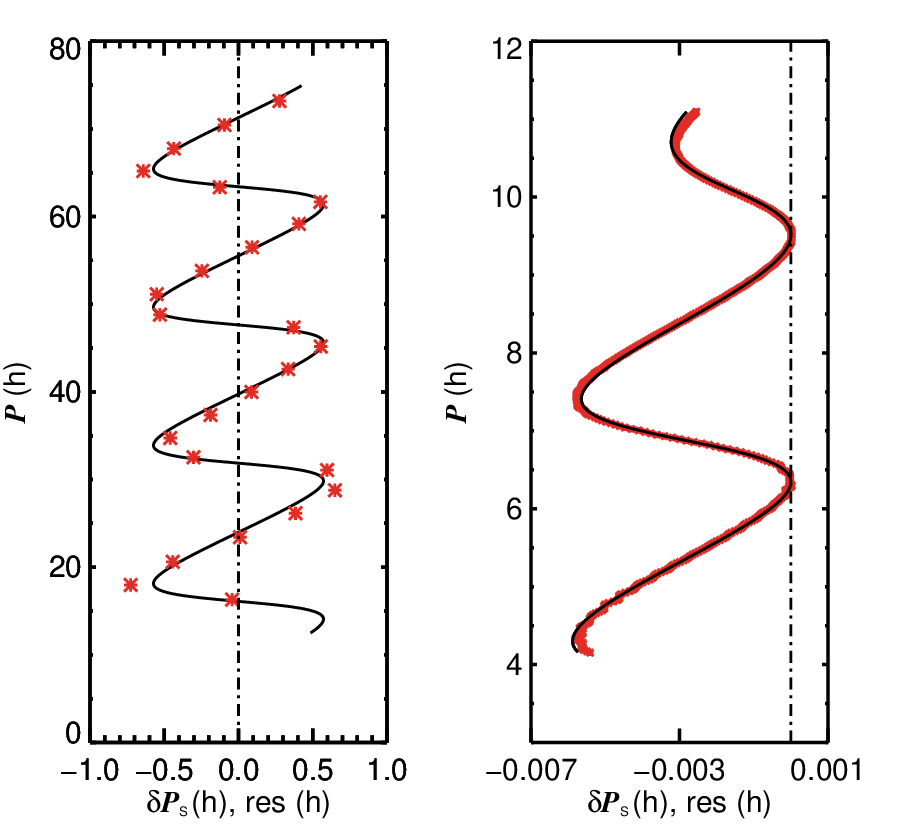}
   \caption{Period échelle diagrams representing the pulsations in the \small{ASTEC} models (red asteriscs) and the respective best fits derived from equation~(\ref{period_as}) (black lines). Left panel: main-sequence model and glitch phase given in equation (\ref{phi_minus}). Right panel: RGB model and glitch phase given in equation (\ref{phi_gau}). The vertical dashed-dotted black line marks res=0.}
   
    \label{fig:echelle_MS_RGB}
\end{figure}

\subsection{Limit of small glitch}

In the case of a step-like glitch with amplitude much smaller than 1, the expression for the periods becomes,
{\begin{equation}
                P=P_0-P_0\frac{\Delta P_{\rm as}}{\pi}\left(\int_{r_1}^{r_2}\frac{L\delta N}{2\pi} \frac{\rmd r}{r}\right)+\frac{\Delta P_{\rm as}}{2\pi}A_{\rm st}\sin\left(2\tilde\beta_2\right),
	\label{eq:limit_step}
\end{equation}}
In this small glitch limit $A_{\rm st}=[N_{\rm in}/N_{\rm out}]_{r^\star}-1\approx -\left(\left[N_{\rm out}/N_{\rm in}\right]_{r^\star}-1\right)\approx\left[\Delta N/N_0\right]_{r^\star}$.

Under these conditions, the expression for the period perturbation can also be derived by making use of the variational principle \citep[e.g.][]{gough93}. The derivation is presented in Appendix \ref{apA}\footnote{This limit was also considered in \cite{miglio08}. However, the authors were only concerned with the periodic component of the perturbation and the phase of the periodic component they present differs from the one derived here by $\pi/2$.}, where our choice of reference model was made such that the smooth contribution to the period perturbation (second term on the rhs of equation (\ref{eq:limit_step})) varies with  $(\delta N/N)^2$, becoming negligible for small enough perturbations. Comparison of equations (\ref{eq:step_lim_vp}) and (\ref{eq:limit_step}), confirms that the two derivations provide the same result in this limit, as expected.

\section{Gaussian-like glitch}
\label{sec:gaus}
In the case of the Gaussian-like glitch, \cite{cunha19} modelled the glitch in $N$, taking 
	\begin{equation}
	\frac{\Delta N}{N_0}\approx\frac{A_{\rm G}}{\sqrt{2\pi}\Delta_{\rm g}}\exp{\left(-\frac{(\omega_{\rm g}^r-\omega_{\rm g}^{\star})^2}{2\Delta_{\rm g}^2}\right)}.
	\end{equation}
Given the inherent difficulty in modelling the eigenfunction through the perturbation (\cf discussion in Section~\ref{sec:intro}), it is important to recall that the analysis was performed under the rough assumption that inside the glitch the eigenfunction has the form
\begin{equation}
\Psi\propto K_0^{-1/2}\sin\left(\int K\rmd r+\frac{\pi}{4}\right),
\end{equation}
and that an ad hoc adjustment was made to the exponential function defining the decay of the glitch signature, so as to recover the decay found when using the variational principle to treat the small glitch limit. 
The glitch-induced phase proposed by the authors under the above conditions is given by
\begin{equation}
    \Phi_{\rm G}={\rm arccot}\left[\frac{1}{A_{\rm G}{\it f}_{\omega}^{\Delta_{\rm g}}\sin^2\beta_2}-\cot\beta_2\right],
        \label{phi_gau}
\end{equation}
where, $f_{\omega}^{\Delta_{\rm g}}=\omega^{-1}{\exp}(-2\Delta_g^2\omega^{-2})$.


\subsection{Tests on model data}

{To test the expression for the periods in the presence of a Gaussian-like glitch and compare with the fits to the period spacings performed by \cite{cunha19}, we follow again their example and fit equation (\ref{period_as}) to the pure g-mode periods computed by the authors for a 1~M$_\odot$ red giant branch (RGB) model. The model was computed with the evolution code {\small ASTEC} \citep{jcd08a} and the pure g-mode pulsations were computed by artificially disregarding the p-mode cavity with the ASTER code \cite[see][for details]{cunha15,cunha19}. Here, the glitch (illustrated in their figure 1, right panel) results from the strong chemical gradient that is built at the first dredged-up and that crosses the g-mode cavity during the luminosity bump.  } 

The same procedure as the one described in section \ref{sec:step_tests} was used. For a Gaussian-like glitch there is an additional free parameter, as we must fit as well the width $\Delta_{\rm g}$. The argument of the arc-cotangent for Gaussian glitches crosses 0, instead of jumping from $+\infty$ to $-\infty$. Thus the codomain was chosen to be [0,$\pi$] to avoid a discontinuity in this region. 

The stellar model used to test the Gaussian-like glitch signature has a denser period spectra than the main-sequence model used to test the step-like glitch signature. As a consequence, the data set to be fitted was larger in the Gaussian-like glitch case and not as many iterations were needed. We ran the MCMC for 4000 iterations with a burn-in of 1000, since the distributions took longer to converge to their stationary state. As before, 50 walkers were used to explore the parameter space. {The priors on the parameters were uniform within the intervals provided in Table~\ref{tab:prior} and the respective marginalised distributions for the parameters are shown in Fig.~\ref{fig:corner_gaussian}.}
%

   

The right panel of figure~\ref{fig:echelle_MS_RGB} shows the pure g-mode periods of the \textsc{ASTEC} RGB model in the form of and échelle diagram, where the model periods (red symbols) are plotted against the residuals computed as in Section~\ref{sec:step_tests}.  The period perturbations derived from the fit of equations~(\ref{period_as}) and (\ref{phi_gau}) to the model periods are also shown (black line). Unlike in the case of the step-like glitch, here the perturbation is generally negative (in the case of the fit it is always negative and in the case of the model data it takes small positive values with a maximum of $\sim 3\times 10^{-5}$~h only at very few points). The reason is that $\Phi_{\rm G}$ includes a smooth, frequency-dependent component [more easily identified in the expression for the small glitch limit given in equation~(\ref{phi_gau_vp})] that contributes negatively to the period perturbation $\delta P_{\rm s}$.

The parameters inferred from the fit are given in Table~\ref{tab:parameters_gau}. 
Comparison with the results found by \cite{cunha19} from fitting the period spacings shows agreement within {1-$\sigma$ on the inferred glitch amplitude, glitch width, and period spacings. The phases and glitch positions also show a reasonable agreement ($\sim$~2-$\sigma$)}. Nevertheless, in this case a large difference is seen when comparing the glitch properties inferred from the fittings of the analytical expression and from the model buoyancy frequency (with the amplitude from the fit being more than 1.5~times larger than that derived from the buoyancy frequency). This discrepancy, also found by \cite{cunha19} when fitting the period spacings, will be discussed further in subsection \ref{accuracy}.

\begin{table*}
	\caption{{Parameters derived from the fit of the analytical expression
			for the Gaussian-like glitch [equations~(\ref{period_as}) and (\ref{phi_gau})]  to the 
			period spacing derived from ASTER for the RGB model. The values shown correspond to the median of the distributions and the $68\%$ confidence intervals.  For comparison, the values of the glitch parameters and asymptotic period spacing inferred by fitting the period spacings, as well as the glitch parameters estimated directly from the ASTEC buoyancy frequency profile, are also shown.}}
	\label{tab:parameters_gau}
	\begin{minipage}{0.99\textwidth}
		\resizebox{\linewidth}{!}{%
			\begin{tabular}{|c|c|c|c|c|c|c|c|}
				\hline
				&$P_{\rm s,min}$&$\Delta P_{\rm as}$ (s) & $A_{\rm G}$ ($10^{-6}$ rad/s)  &
				$\omega_{\rm g}^\star$ ($10^{-6}$ rad/s)  & $\Delta_{\rm g}$  ($10^{-6}$ rad/s) &
				$\delta$& reference\\
				\hline
				& & & & & & & \\
          				{Periods' fit}& {$14967.183^{ +0.076}_{-0.074}$}  &{$67.53443^{ +0.00034}_{-0.00035}$} & {$589.1^{ +4.4}_{-4.3}$} & 
				{$1730.6^{ +1.6}_{-1.6}$} &{$157.31^{ +0.55}_{-0.55} $}&  {$-0.7936^{ +0.0069}_{-0.0069}$}& This work\\
				& & & & & & &\\
				{Period spacings' fit}&--&{$67.534^{ +0.005}_{-0.005}$} & {$607^{ +27}_{-25}$} & 
				{$1747.3^{ +7.6}_{-7.7}$} &{$158.5^{ +3.4}_{-3.4} $}&  {$-0.872^{ +0.034}_{-0.034}$}& \cite{cunha19}\\
				& & & & & & &\\
				\hline
				{Estimated}&--&--&{$380$} & {$1632$} & {$156$} &--& \cite{cunha19}\\
				\hline
			\end{tabular}
		}
	\end{minipage}
\end{table*}

\subsection{Dirac $\delta$ limit}
\label{dirac}
To verify that the phase for the Gaussian-like glitch reproduces the expected form in the limit of a Dirac $\delta$, we consider  equation~(\ref{phi_gau}) in that limit and compare with the phase derived by \cite{cunha15}.
Recalling that
	\begin{equation}
	\lim_{\Delta_g\to 0^+}\frac{1}{\sqrt{2\pi}\Delta_{\rm g}}\exp{\left(-\frac{(\omega_{\rm g}^r-\omega_{\rm g}^{\star})^2}{2\Delta_{\rm g}^2}\right)} = \delta(\omega_{\rm g}^r-\omega_{\rm g}^{\star}),
 \label{eq:GtoDir}
	\end{equation}
 we find that the glitch-induced phase tends to
\begin{equation}
    \Phi_{\rm G,lim}={\rm arccot}\left[\frac{1}{A_{\rm G}\omega^{-1}\sin^2\beta_2}-\cot\beta_2\right],
        \label{phi_gau_dirac}
\end{equation}
 in the limit when the glitch tends to a Dirac $\delta$ in the variable $\omega_{\rm g}^r$.
 
 In \cite{cunha15}, the authors defined the glitch in $N^2(r)$ as,
    \begin{equation}
	\frac{\Delta N^2}{N_0^2}\approx A\delta(r-r^\star).
	\end{equation}
By integrating the wave equation (\ref{waveeqap}) once across the glitch, the authors found the discontinuity in the radial derivative of the eigenfunction to be given by (see their equation 11)
    \begin{eqnarray}
    \left[\Psi_{\rm out}^\prime-\Psi_{\rm in}^\prime\right]_{r^\star}=-\lim_{\Delta r\to 0}\int_{r^\star-\Delta r}^{r^\star+\Delta r}\Delta K ^2\Psi\rmd r= \nonumber \\ -\lim_{\Delta r\to 0}\int_{r^\star-\Delta r}^{r^\star+\Delta r}K_0 ^2A\delta\left(r-r^\star\right)\Psi\rmd r.
	\end{eqnarray}
Considering the same discontinuity in the case of the Gaussian-like glitch \cite[][see their Appendix A1]{cunha19} and taking the Dirac $\delta$ limit, one has, in terms of the variable $\omega_g^r$,
    \begin{eqnarray}
    \left[\Psi_{\rm out}^\prime-\Psi_{\rm in}^\prime\right]_{r^\star}= -\lim_{\Delta\omega_g^r\to 0}\int_{\omega_g^\star-\Delta\omega_g^r}^{\omega_g^\star+\Delta\omega_g^r}{\omega}^{-1}\Delta K \Psi{\rmd\omega_g^r}=  \nonumber \\ -\lim_{\Delta\omega_g^r\to 0}\int_{\omega_g^\star-\Delta\omega_g^r}^{\omega_g^\star+\Delta\omega_g^r}{\omega}^{-1}K_0 A_G\delta(\omega_g^r-\omega_g^{r^\star})\Psi \rmd\omega_g^r,
	\end{eqnarray}
where we have used (\ref{eq:GtoDir}) to convert from the Gaussian perturbation to the corresponding Dirac $\delta$ limit. We thus find $A_G=(LN_0^\star/r^\star) A$ in the Dirac $\delta$ limit. Substituting in equation~(\ref{phi_gau_dirac}), we confirm that the Gaussian glitch-induced phase is the same as that derived by \cite{cunha15} (\cf their equation 16) in the Dirac $\delta$ limit.

\subsection{Limit of small glitch}
\label{small_G}
When $\Delta N/N_0 \ll 1$ everywhere, $A_G/\omega \ll 1$ (since $\Delta_g <\sim\omega$ for a glitch to exist). In that case,
{\begin{equation}
          \Phi_{\rm G}\sim\arctan\left[{A_{\rm G}{\it f}_{\omega}^{\Delta_{\rm g}}\sin^2\beta_2}\right]\sim\frac{1}{2}A_{\rm G}{\it f}_{\omega}^{\Delta_{\rm g}}\left[1-\cos\left(2\beta_2\right)\right],
       \label{phi_gau_vp}
\end{equation}}
 Defining the reference model as that with an otherwise equal buoyancy frequency, without the Gaussian perturbation, we have that $\delta N/N_0=\Delta N/N_0$. Accordingly, the period perturbations are derived from equation~(\ref{period_as}) to be,
{  \begin{equation}
                P=P_0-P_0\frac{\Delta P_{\rm as}}{2\pi^2}{A_G}-\frac{\Delta P_{\rm as}}{2\pi}{A_{G}}{\it f}_\omega^{\Delta_g}\left[1-\cos\left(2\beta_2\right)\right].
	\label{eq:limit_gaus}
\end{equation}}
Under this limit of a small perturbation to the buoyancy frequency, the perturbation to the periods induced by a Gaussian-like glitch can be derived with recourse to the variational principle. The derivation is presented in Appendix~\ref{apA} with the period perturbations given by equation~(\ref{eq:gaus_lim_vp}). Comparison with equation~(\ref{eq:limit_gaus}) shows that the sinusoidal part of the glitch-induced period perturbation is a factor of two smaller in the case derived by matching the eigenfunctions through the approach followed by \cite{cunha19}. Moreover, the smooth component resulting from this derivation has also an additional term compared to the variational approach, which is generally not negligible. In practice, the difference in the smooth component will have no impact  when inferring the glitch properties by fitting the periods or period spacings (at most, it will introduce small differences in the inferred unperturbed periods that are of no physical interest since they refer to a hypothetical unperturbed model). However, the factor of two in the sinusoidal component will be translated into a factor of two in the inferred amplitude of the glitch, which will be two times larger when fitting the expression derived through the approach followed by \cite{cunha19}, compared with fitting with the expression derived from the variational principle. Changing $\Phi_{\rm G}$ such that the perturbation to the periods satisfies that derived through the variational principle in the limit of a small glitch is not an option because we verified that the expression for $\Phi_{\rm G}$ satisfies the correct limit when the glitch approaches a Dirac $\delta$  and the inherent difficulty of fitting the eigenfunctions at the perturbation disappears (Section~\ref{dirac}).  Instead, in Section~\ref{accuracy} we explore the accuracy of the analytical expression for $\Phi_{\rm G}$ in intermediate cases, simulating glitches of different amplitudes and widths, and estimating the error incurred on the inferred glitch parameters.

\subsection{Tests to the accuracy of the Gaussian-like glitch inferences}
 \label{accuracy}
The results from Section~\ref{small_G} illustrate well the anticipated difficulty in deriving an expression for the glitch-induced signature that is valid across the parameter space that one would like to explore, in the case of a Gaussian-like glitch. It is thus important to test the accuracy of the proposed expression on control data before using it to fit real data. To that end, we simulated a series of Gaussian-like glitches, starting from the buoyancy frequency of our 1M$_\odot$ red giant model. Specifically, we started by building a glitchless reference model by either removing or smoothing the features in $N$ that produce visible period spacing variations on scales smaller than the frequency range analysed (\ie [25,67]$~\mu$Hz). Two features were identified: (1) the glitch in $N$ associated to the steep chemical gradient left by the first dredge-up, which was removed by fitting a second order polynomial across the glitch region; (2) a series of significant short-scale variations in the second derivative of $N$ (of numerical origin) taking place in the H-burning shell, which were suppressed by applying a boxcar average to $N$ (with a width of $\sim$1/50 of the shell) in the region of the H-burning shell. The buoyancy frequency of the original \textsc{ASTEC} model (dashed black line) around the glitch is compared with that of the glitchless reference (black dotted line) in Fig.~\ref{fig:reference}.

\begin{figure}
	\includegraphics[width=\columnwidth]{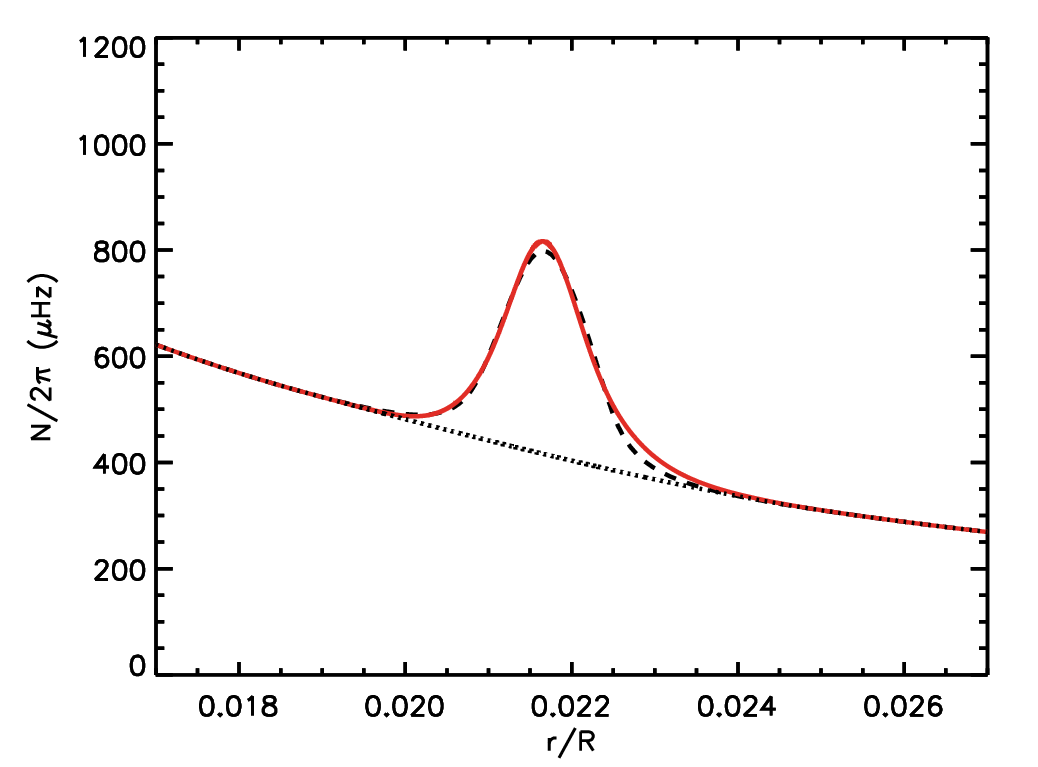}
   \caption{Buoyancy frequency for the original \textsc{ASTEC} model (dashed black line), the glitchless reference model (dotted black line), and model 5, built by adding to the glitchless reference model a glitch with properties similar to that of the original \textsc{ASTEC} model.}
    \label{fig:reference}
\end{figure}

\begin{figure}
	\includegraphics[width=\columnwidth]{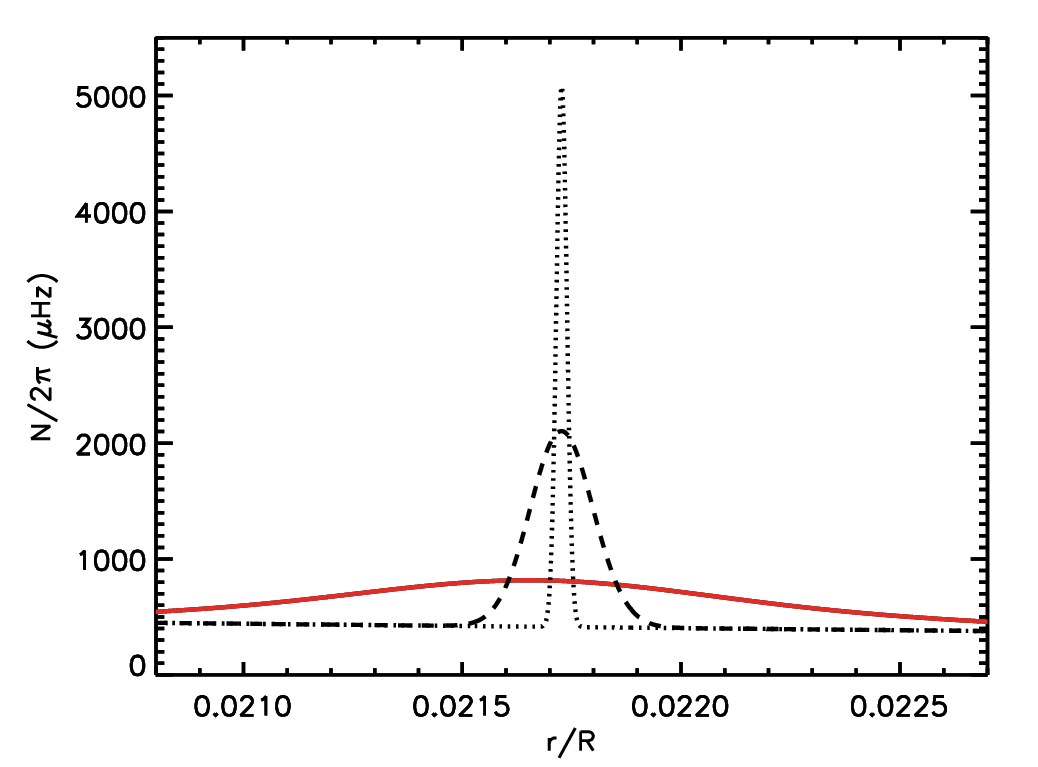}
    \includegraphics[width=\columnwidth]{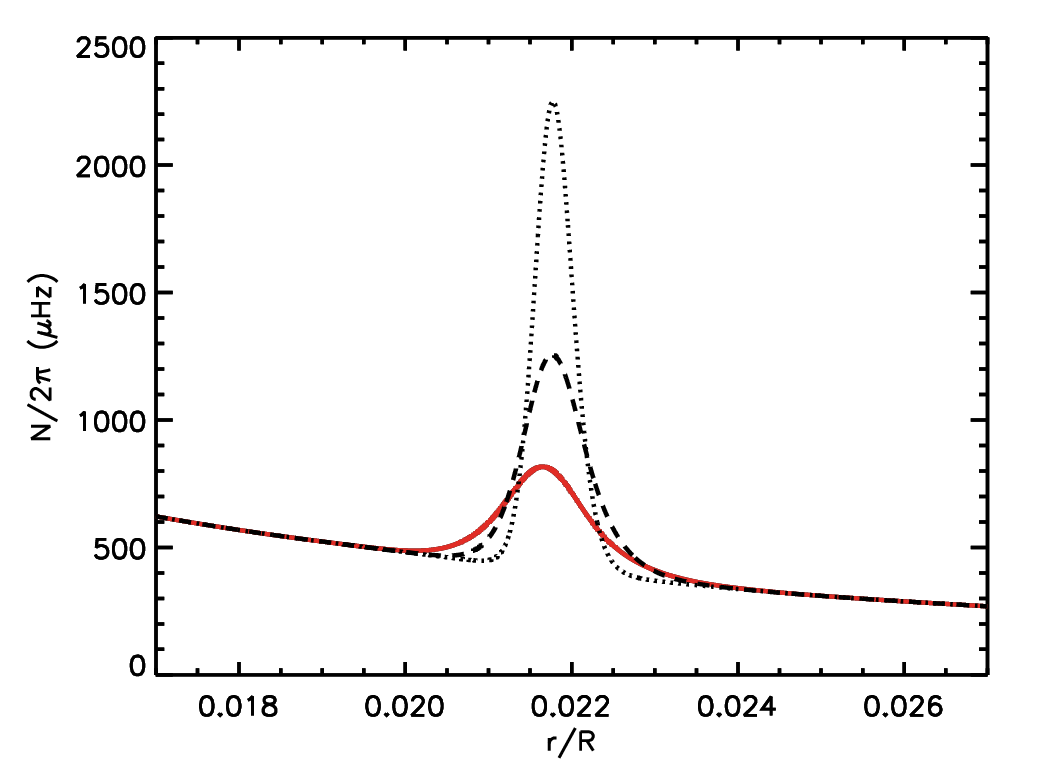}
    \includegraphics[width=\columnwidth]{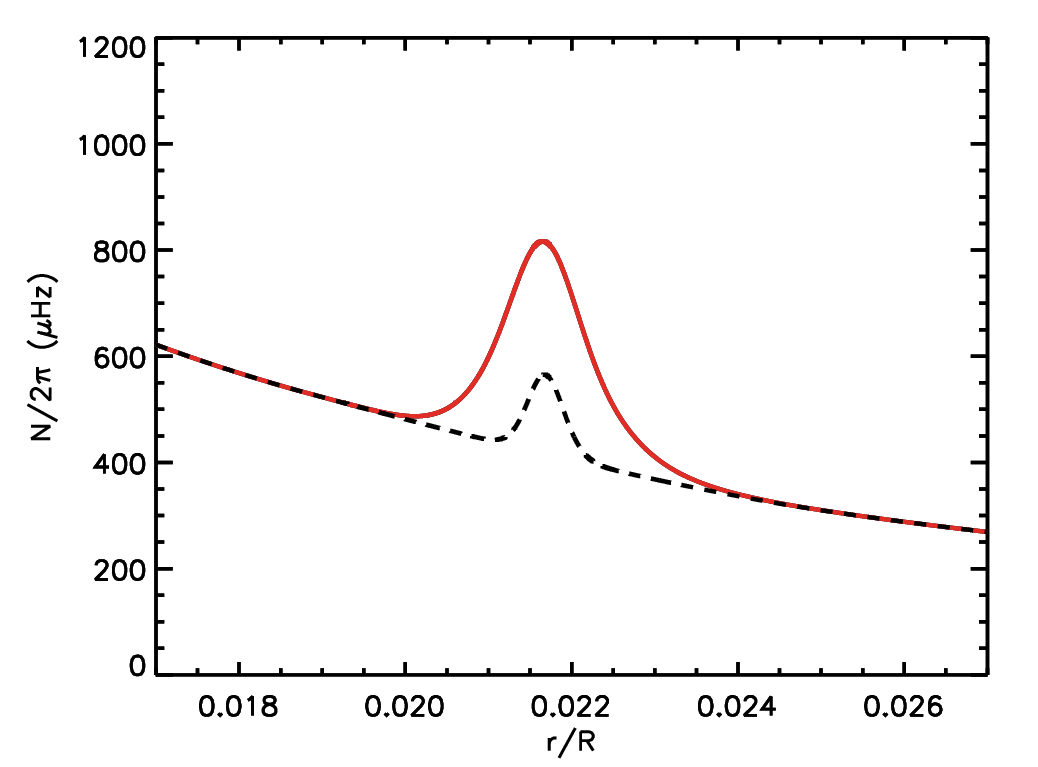}
   \caption{Buoyancy frequency for the glitch-simulated models near the glitch position. In all panels, the continuous red  line shows the case of model 5, corresponding to a glitch with properties similar to the original \textsc{ASTEC} model. Additionally, the black lines are as follows. Top panel: Models 1 (dotted line) and 2 (dashed line). Middle panel: Models 3 (dotted line) and 4 (dashed line). Bottom panel: Model 6.  }
    \label{fig:brunt_simulated}
\end{figure}

\begin{figure}
	\includegraphics[width=\columnwidth]{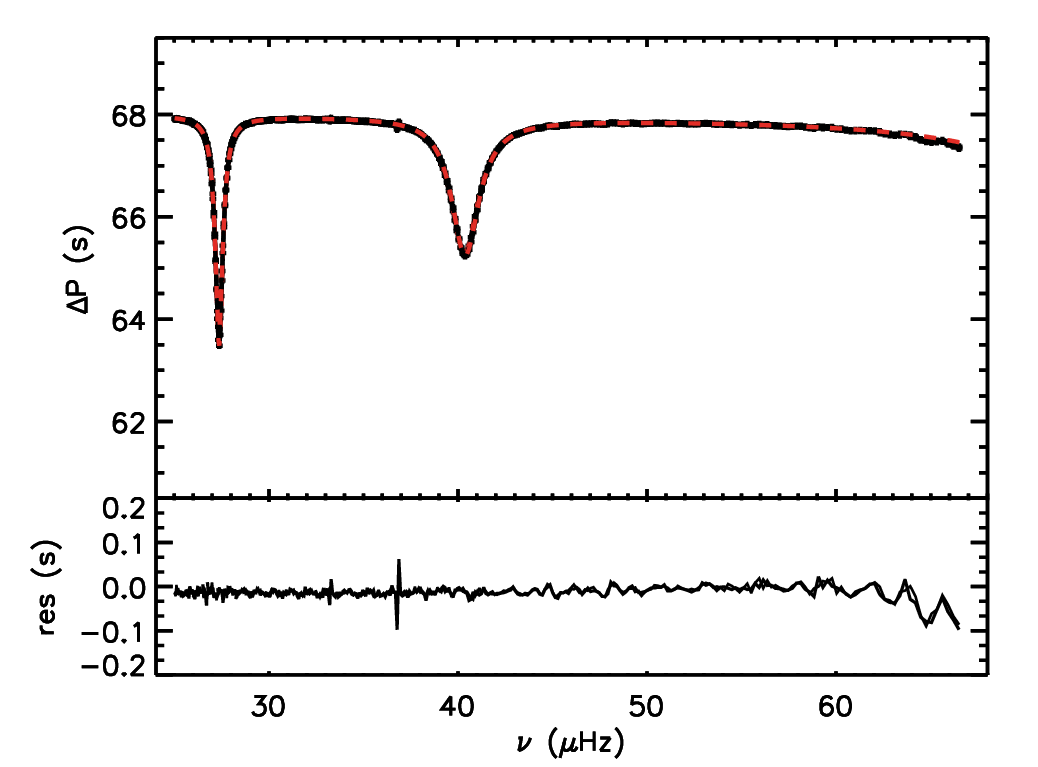}
 	\includegraphics[width=\columnwidth]{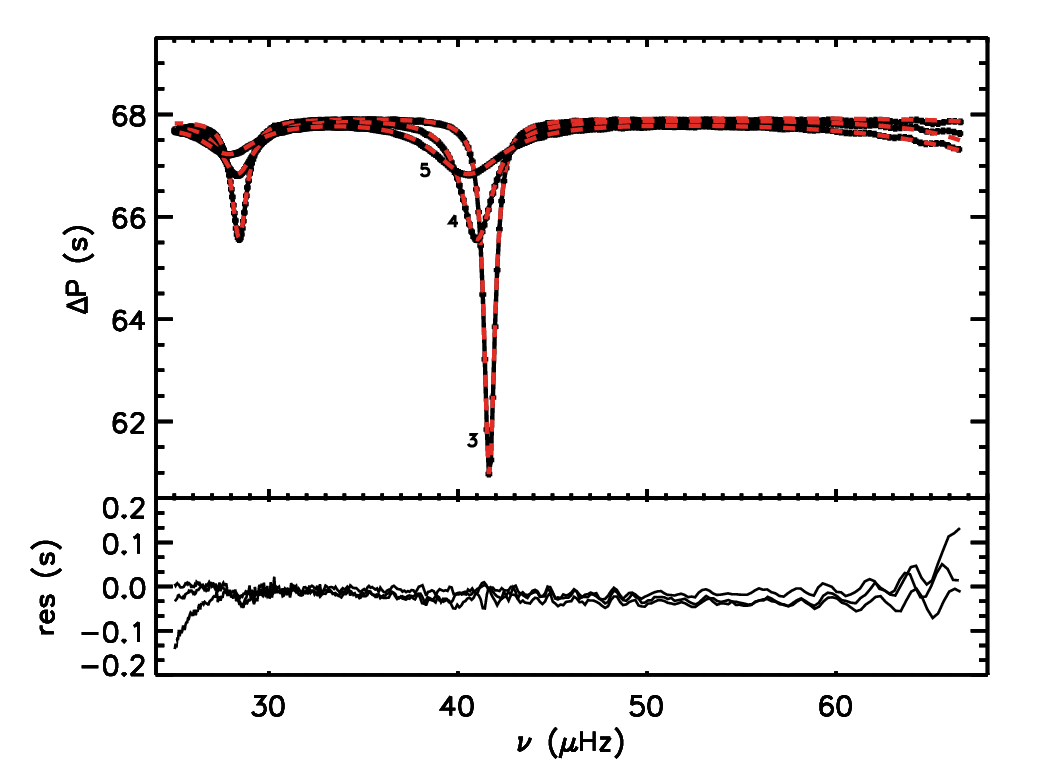}
	\includegraphics[width=\columnwidth]{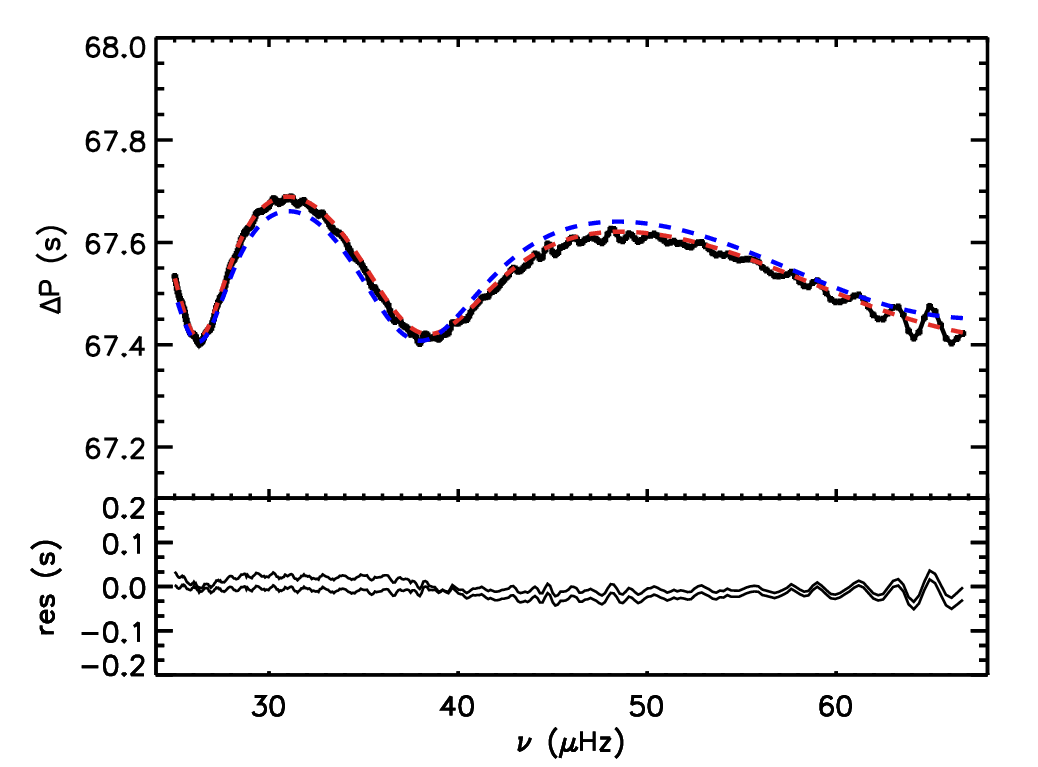}
   \caption{Period spacings from \textsc{ASTER} for the simulated Gaussian-like glitches  listed in Table~\ref{tab:simulated} (black line and plus symbols) and respective best analytical-model fits  (dashed red line). The residuals (\textsc{ASTER} -- best fit) are also shown for each case. Top: models 1 and 2. Middle: models 3, 4 and 5. Bottom: model 6. The fits in red included a linear smooth component. The dashed blue line in the bottom panel shows the fit without the inclusion of a smooth component. For all other models the fits with and without the inclusion of the smooth component would be indistinguishable in these figures.}
    \label{fig:simulated}
\end{figure}

Next we artificially added one Gaussian glitch at a time to the glitchless reference model, to generate a set of six models, each with a single glitch. The parameters of the added glitches were first defined based on the buoyancy depth of the reference model. As $N$ is modified by the addition of the glitch, we then recomputed the buoyancy depth based on the new $N$ and fitted a Gaussian function to the glitch to extract the true glitch parameters  that are to be compared with the values inferred from fitting the analytical expression. The parameters of the six artificial glitches were chosen to explore different regimes, namely: (1) three glitches with similar widths and different amplitudes were simulated to investigate the impact of increasing amplitude; (2) two glitches with similar amplitude and different small widths were simulated to verify the impact of approaching the Dirac $\delta$ limit; (3) one glitch with a small amplitude was also simulated to test the results when approaching the small glitch limit. This series of glitches is illustrated in Fig.~\ref{fig:brunt_simulated} and the respective properties are provided in Table~\ref{tab:simulated}. We note that one of these (model 5) has properties similar to those of the original \textsc{ASTEC} model and is shown in red in Fig.~\ref{fig:reference} and all panels of Fig.~\ref{fig:brunt_simulated}, for comparison.

\begin{table}
	\caption{Properties of the simulated glitches used in Section~\ref{accuracy}.}
	\label{tab:simulated}
	\begin{minipage}{0.48\textwidth}
		\resizebox{\linewidth}{!}{%
			\begin{tabular}{|c|c|c|c|c|}
   \hline									
Model ID	&	$A_{\rm G}$ ($10^{-6}$ rad/s)	&	$\omega_{\rm g}^\star$ ($10^{-6}$ rad/s)	&	$\Delta_{\rm g}$ 	&	Comment	\\
\hline									
1	&	502.2	&	1570.1	&	15.5	&	Dirac-like	\\
2	&	493.9	&	1603.7	&	43.6	&		\\
3	&	1920.7	&	1752.7	&	154.8	&	Largest amplitude	\\
4	&	837.6	&	1700.4	&	154.8	&		\\
5	&	355.9	&	1658.1	&	147.1	&	Similar to original	\\
6	&	43.0	&	1565.0	&	46.9	&	Small	\\
\hline									
			\end{tabular}
		}
	\end{minipage}
\end{table}

The pure g-mode periods of the simulated glitch models were computed with the code \textsc{ASTER} and the respective period spacings were fitted with the analytical expression proposed by \cite{cunha19} (their equation 15) following the procedure described in that  paper. In the case of the smallest glitch (model 6) the fit required adding a smooth component to the analytical expression describing the period spacings, which we modelled as a linear function, $S_{\rm m} = a_0 +a_1\omega$, where $a_0$ and $a_1$ are constants. We thus added the smooth component to the fit of the period spacings of all simulated glitch models for consistency, although no significant impact was found in any of the other cases. 

The best fits to the period spacings for all six simulated glitch models are shown in Fig.~\ref{fig:simulated}, with the residuals shown in the bottom part of each panel. Comparison of the model period spacings (in black) with the best fits (in red) shows that the analytical expression proposed by \cite{cunha19} (their equation 15) for the Gaussian-like glitch signature reproduces well the period spacings computed with {\small ASTER} in all cases (with all deviations smaller than 0.3$\%$). Inspection of the blue curve in the bottom panel also shows that for the smallest  glitch (Model ID 6 in Table~\ref{tab:simulated}) a smooth component needs to be added to the analytical expression in order to achieve a good fit. With the addition of the smooth component, the fit is indeed improved significantly, as seen by the model shown in red in the same panel.

In the top panel of Fig.~\ref{fig:simulated} we can also see how the signature of the glitch changes as the glitch becomes narrow. The two glitch models shown correspond to the buoyancy frequency profiles in black seen in the top panel of Fig.~\ref{fig:brunt_simulated}.  As we approach the Dirac~$\delta$ limit we have seen that the glitch signature becomes independent of the glitch width [\cf equation~(\ref{phi_gau_dirac})]. This is reflected in the results seen in the top panel of Fig.~\ref{fig:simulated}, as the glitch signatures for the two cases are almost indistinguishable (there are two black and two red lines plotted in this panel). Notice that the residuals between the model data (in black) and the fits (in red) for the two cases, also shown in the bottom part of the top panel, are also hardly distinguishable.

Likewise, we fitted the {\small ASTER} periods for the six simulated models with equation (\ref{period_as}), using the glitch phase defined by equation~(\ref{phi_gau}), adding a second order smooth component to mimic the impact of the first order smooth component added to the analytical expression for the period spacings. Figure~\ref{fig:PSvsPeriods} shows the ratios between the inferred and true glitch properties both for the case of the fit to the period spacings and for case of the fit to the periods. 

Inspection of the inferred glitch parameters show that these are not accurate, in the sense that the error bars associated with the inferences are small compared to the distance to the true values of the glitch properties. Thus, while the proposed analytical expressions for the signature on the periods and period spacings of the Gaussian-like glitch reproduce well the model data, they produce systematic errors in the inferred glitch properties. This was to be expected given the difficulties associated with the derivation of the analytical expression in this case (\cf Section~\ref{sec:gaus}). Nevertheless, the tests with artificial glitch models performed here allow us to understand the extent of the systematic errors so that these can be considered in applications to real data. According to the results, the parameter that is most affected, namely the amplitude of the glitch, can still be recovered within a factor of two for all cases tested. In fact, we verify that the maximum error of a factor of two occurs for the smallest amplitude glitch, as already anticipated through the inspection of that limit in Section~\ref{small_G}. On the other hand, we see that the inferred amplitude becomes closer to the true value when the glitch becomes narrower, as expected from the Dirac~$\delta$ limit discussed in Section~\ref{dirac}. Regarding the glitch width, the largest errors occur for the narrower glitch (model ID 1 in Table~\ref{tab:simulated}), where the glitch-induced signature becomes less sensitive on this parameter. For this case we notice also a significant difference in the width inferences made by fitting the periods and the period spacings, and an increase in the error bars, compared to those obtained for the other glitch models. Finally, we see that the position of the glitch is determined within 10$\%$ of the true value in all cases considered.

\begin{figure*}
	\includegraphics[width=2\columnwidth]{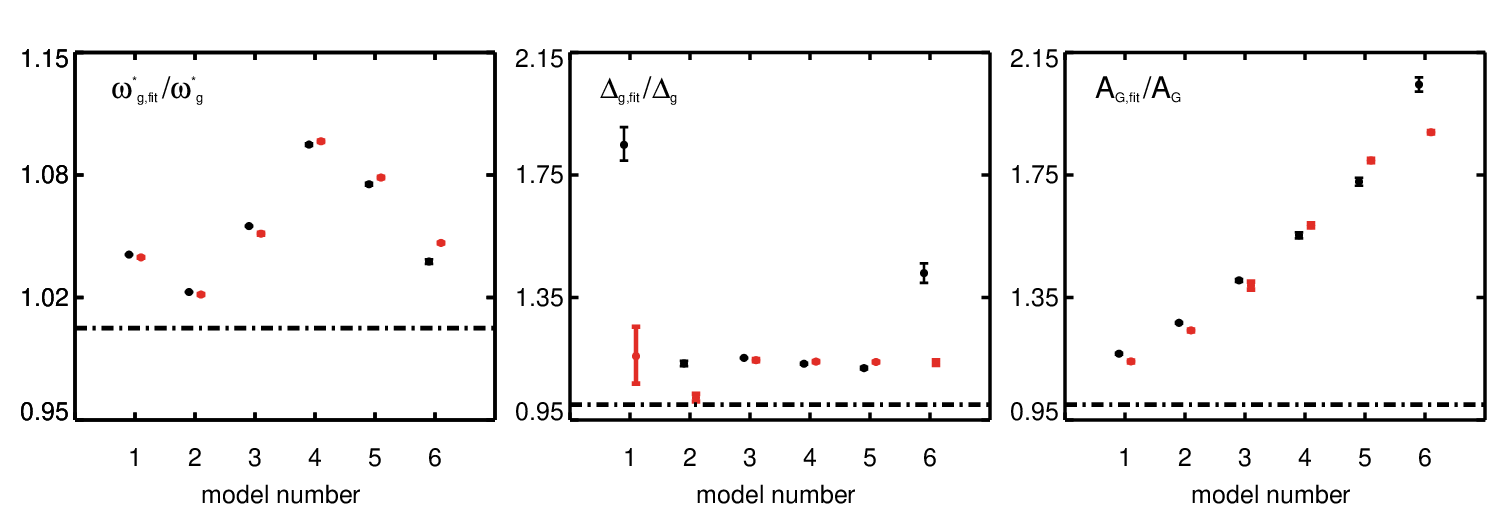}
   \caption{Ratios between the inferred and true glitch parameters. Buoyancy radius (left), glitch width (middle), glitch amplitude (right). A comparison is shown of the values inferred from the fits to the period spacings (black) and to the periods (red). Note that the symbols have been displaced horizontally for a better visualisation. The horizontal dashed-dotted black line represents a ratio equal to 1. 
   }
    \label{fig:PSvsPeriods}
\end{figure*}

\section{Mixed modes}
\label{sec:additional}
For completeness, in this section we discuss the expression for the  frequency perturbations resulting from the impact of buoyancy glitches on mixed modes. 
The perturbation to the mixed-mode period spacings induced by a buoyancy glitch was first considered for the case of a glitch represented by a Dirac-$\delta$ function \citep{cunha15}  and later for a Gaussian-like glitch \citep{cunha19}. 
Following on these works, \cite{vrard22} provided an expression for the frequency perturbation of dipole modes, namely,
\begin{equation}
    \nu=\nu_{{\rm a},n}+\frac{\Delta\nu}{\pi}\arctan\left\{q\tan\left[\pi\left(\frac{1}{\nu\Delta P_{{\rm as}}}-\epsilon_{\rm g}\right)+\Phi\right]\right\},
    \label{eq:mixed}
\end{equation}
where $\nu_{{\rm a},n}$ represents the frequency of the pure acoustic dipole mode of (pressure) radial order $n$ that would exist in the absent of mode coupling, $\Delta\nu$ is the large frequency separation, $q$ is the coupling factor and $\epsilon_{\rm g}$ is the gravity phase offset. This expression assumes that the integral of $N/r$ within the g-mode cavity is the same in the
glitch and glitchless (reference) models. The impact of the glitch is incorporated in glitch phase $\Phi$, as before. 

Explicit forms for $\Phi$ were derived by \cite{cunha15} and \cite{cunha19} for the Dirac-$\delta$ and Gaussian-like glitch, respectively. In both cases the authors assumed that the glitch was located in the outer half of the g-mode cavity. {However, from the point of view of the solutions within the g-mode cavity, we can think of this case as being similar to the one of pure g modes, with the exception that the boundary conditions applied beyond the two turning points are no longer the same, as a result of the coupling outwardly from the cavity. In other words,} the presence of the p-mode cavity outwards from the g-mode cavity breaks the symmetry of the problem with respect to the solution inside the g-mode cavity.  It is thus worth considering how the glitch phase is modified when considering a glitch in the inner half of the g-mode cavity. That analysis is presented in  Appendix~\ref{apC}.

In all cases the functional form of $\Phi$ for mixed modes is found to be unchanged with respect to that derived in the analysis for the pure g modes, that is, it is given by equations~(\ref{phi_minus}), (\ref{phi_gau}), and (\ref{phi_gau_dirac}), for the step-like, Gaussian-like, and Dirac-$\delta$ glitch, respectively. 
 However, while the definition of $\tilde\beta_2$, used in the case of a glitch in the inner half of the g-mode cavity, remains the same as in the pure g-mode case, the definition of $\beta_2$ used in the case of a glitch in the outer half of the g-mode cavity differs from that derived in the pure g-mode case, as found in the previous works \citep{cunha15,cunha19}. In particular, when employing any of the expressions for the phase $\Phi$ for mixed modes with a glitch in the outer half of the g-mode cavity, one must make the substitution  
 { $\beta_2\rightarrow\beta_{2,\varphi} \approx \frac{\omega_{\rm g}^\star}{2\pi} P+\pi/4+\delta +\varphi$, where $\varphi$ is the coupling phase defined by}
\begin{equation}
\varphi\approx{\arctan}\left[\frac{q}{\tan\left[\pi\left(\nu-\nu_{{\rm a},n}\right)/\Delta\nu\right]}\right].
\label{varphi}
\end{equation}
As a result of the inclusion (exclusion) of $\varphi$ in the arguments of the sinusoidal functions entering the definition of $\Phi$ in the case of glitches in the outer (inner) half of the g-mode cavity, different glitch signatures on the frequencies of mixed modes are expected depending on which case is considered. A thorough discussion on this symmetry breaking will be presented in a separate work.  

Finally, we emphasise also the non-additive nature of the effects of the glitch and coupling on the frequencies, as already discussed in \cite{cunha19} (their section 2). This requires that the two effects are considered simultaneously, by fitting with expression (\ref{eq:mixed}). {While streching of the periods can be performed \citep{mosser15,ong23}, caution must be exerted on the interpretation of the results from that streching, particularly if the frequencies are observed only near the frequency of maximum coupling. }

Fits of expression (\ref{eq:mixed}) to Kepler data on a sample of helium core-burning stars with a glitch located in the inner half of the g-mode cavity have recently been presented by \cite{vrard22}, enabling the first characterisation of core glitches in these stars.

\section{Conclusions}
\label{sec:conclusions}
In this work we provided and tested on model data analytical expressions for the perturbations induced by a glitch located in the g-mode cavity of a star on pure g-mode periods and mixed-mode frequencies. This complements the work published previously on the period-spacings perturbations \citep{cunha15,cunha19} and will be particularly useful when fitting data for which the observed periods are not sequential in radial order {(where the computation of the period spacings is not accessible for all radial orders, but an estimate of the asymptotic period spacing is still available}). 

Using model data, we have also tested the accuracy of the analytical expression proposed for the seismic signature of Gaussian-like glitches, where drastic approximations had to be made in the derivation due to the finite width of the glitch. We found that the analytical expression for the periods and period-spacings perturbations becomes insensitive to the glitch width when the glitch approached the Dirac-$\delta$ limit. In that case, the recovery of the glitch width becomes less accurate. Nevertheless, for all cases studied here, we found that the parameter that is most affect, namely the glitch amplitude, can still be recovered from the fitting within a factor of two. 

{Finally, comparing the glitch-induced phases for step-like and Gaussian-like glitches, we see that glitch signatures have different dependencies on period (see Fig.~\ref{fig:echelle_MS_RGB}), according to their shapes. This introduces an additional uncertainty in the interpretation of the glitch amplitude (and width), particularly when the range of observed periods does not cover a significant number of cycles of the glitch signature. Therefore, we are forced to conclude that the glitch position is generally the most robust of the inferred glitch parameters, similar to what is found in studies of acoustic glitches \cite[e.g.][]{mazumdar14}.}


\begin{acknowledgements}
The authors would like to thank the anonymous referee for the interesting discussion and suggestions made during the refereeing process. This work has been supported by Fundação para a Ciência e Tecnologia FCT-MCTES, Portugal, through national funds by these grants UIDB/04434/2020 (DOI: 10.54499/UIDB/04434/2020), UIDP/04434/2020.FCT (DOI: 10.54499/UIDP/04434/2020) and 2022.03993.PTDC (DOI:10.54499/2022.03993.PTDC). MC is funded by FCT-MCTES by the contract with reference CEECIND/02619/2017. 

\end{acknowledgements}




\bibliographystyle{aa}
\bibliography{solar-like} 




\begin{appendix}
\section{Variational principle}
\label{apA}
Our starting point is the wave equation A1 of Appendix A of \cite{cunha19} written in terms of the variable $\Psi=~(r^3/g\rho \tilde f) ^{1/2}\delta p$ (\cf section~\ref{sec:periods}). The equation resulting from the linear, adiabatic pulsation equations for the case of a spherically symmetric equilibrium
under the Cowling approximation, is then
\begin{eqnarray}
{\Psi^{\prime\prime}}+K^2\Psi=0,
\label{waveeqap1}
\end{eqnarray}
where a prime represents  a differentiation with respect to $r$ and we shall approximate the radial wavenumber $K$ as in equation~(\ref{k2}), given our interest in pure g modes.

Multiplying equation (\ref{waveeqap1}) by the complex conjugate $\Psi^\dagger$, integrating once by parts between the turning points $r_1$ and $r_2$ and rearranging, we find the following integral equation for the periods:
\begin{equation}
    P^2=\frac{\int_{r_1}^{r_2}(\Psi^\prime)^2\rmd r+\int_{r_1}^{r_2}\frac{L^2}{r^2}\Psi^2\rmd r-\left[\Psi^\prime\Psi^\dagger\right]_{r_1}^{r_2}}{\int_{r_1}^{r_2}\frac{L^2N^2}{4\pi^2 r^2}\Psi^2\rmd r}\equiv\frac{F}{I}.
\end{equation}

Next we consider a small localised perturbation to a reference model characterised by $N_0$, such that no change is induced to the turning points. Small perturbations to the periods induced by the change to the reference model are given by
\begin{equation}
\delta P^2 \approx \frac{\delta F-P^2\delta I}{I_0}.
\end{equation}
To first order, the perturbation of the eigenfunctions do not contribute to the perturbation of the periods. Thus $\delta F=0$ and we find
\begin{equation}
    \delta P\approx -\frac{P}{2}\frac{\delta I}{I_0}=-\frac{P}{2}\frac{\int_{r_1}^{r_2}\frac{L^2\delta N^2}{4\pi^2r^2}{\Psi_0^2\rmd r}}{\int_{r_1}^{r_2}\frac{L^2 N_0^2}{4\pi^2r^2}{\Psi_0^2\rmd r}},
    \label{eq:periodpert_ap}
\end{equation}
where $\Psi_0$ is the solution in the reference model and is asymptotically given by \cite{gough93},
\begin{equation}
    \Psi_{0} \sim \tilde\Psi_{0} K_{0}^{-1/2}
\sin\left(\int_{r_1}^r K_{0}\rmd r +
  \frac{\pi}{4}\right),
\label{gasymp}
\end{equation}
and $\tilde\Psi_0$ is a constant.
The integral in the denominator of equation~(\ref{eq:periodpert_ap}) can be readily calculated. Taking $K_0\approx LN_0/r\omega$ in the amplitude of the eigenfuntion $\Psi_0$, we find
\begin{equation}
    I_0\approx\frac{\omega}{4\pi^2}\tilde\Psi_0^2\int_{0}^{\tilde\omega_{\rm g,0}}\sin^2\left(\frac{\tilde\omega_{\rm g,0}^r}{\omega}+\frac{\pi}{4}+\tilde\delta\right)\rmd\tilde\omega_{\rm g,0}^r\approx\frac{\omega}{8\pi^2}\tilde\Psi_0^2\omega_{\rm g,0}.
    \label{I0}
\end{equation}

\subsection{Small step-like glitch}
\label{apa_st}

To proceed with the calculation of $\delta P$ we need first to define the reference model and the glitch shape. Here we consider the case of a decreasing step-like glitch in the inner half of the g-mode cavity and define the reference model such that to first order in $\delta N/N_0$ there is no smooth contribution to the period perturbations. Specifically, we define a linearly varying $N_0$ around the glitch given by
\begin{equation}
    N_0=N_{\rm in}+\frac{\Delta N}{\Delta\omega_{\rm g}}\left(\tilde\omega_{\rm g}^{\rm a}-\tilde\omega_{\rm g}^{\rm r}\right),
\end{equation}
where $\Delta\omega_{\rm g}=\tilde\omega_{\rm g}^{\rm b}-\tilde\omega_{\rm g}^{\rm a}$ and $\Delta N=N_{\rm in}-N_{\rm out}$. Here, $\tilde\omega_{\rm g}^{\rm a}$ and $\tilde\omega_{\rm g}^{\rm b}$ are equidistant from $\tilde\omega_{\rm g}^{\star}$ and 
we assume the transition of $N_0$ to $N_{\rm in}$ and $N_{\rm out}$ on each side of $\tilde\omega_{\rm g}^{\star}$ occurs on a scale much greater than the local characteristic scale of the wave, ensuring that $\Delta\omega_{\rm g}\gg \omega$. Also, any variations of $N$ within the interval $\Delta\omega_{\rm g}$ in addition to the jump at $\tilde\omega_{\rm g}^\star$ are ignored, by comparison with the variation at the glitch. A schematic view of $N_0$ is shown in Fig.~\ref{fig:scheeme}.
\begin{figure}
	\includegraphics[width=0.9\columnwidth]{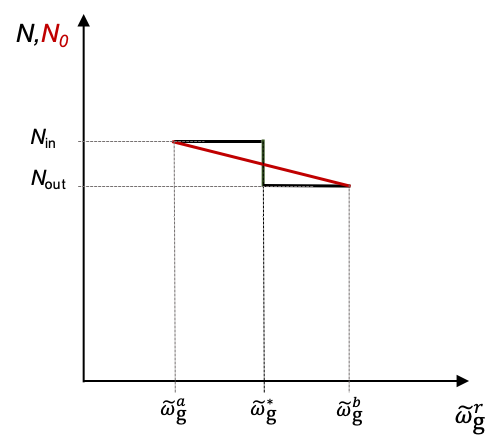}
   \caption{Schematic view of the buoyancy frequency in the reference model around the glitch ($N_0$, in red) in comparison to that in the model with the glitch ($N$, in black). Quantities are shown in arbitrary units.}
    \label{fig:scheeme}
\end{figure}
While this is clearly a simplified reference model, in particular because its first derivative is discontinuous at the points $\tilde\omega_{\rm g}^{\rm a}$ and $\tilde\omega_{\rm g}^{\rm b}$ where it merges into $N$, it suffices our purpose, as the dominant feature in the perturbation $\delta N$ defined from this reference model is still the glitch in $N$. It is clear from symmetry, that in this case
\begin{equation}
    \int_{0}^{\tilde\omega_{\rm g}}\delta N\rmd\tilde\omega_{\rm g,0}^r= \int_{r_1}^{r_2}LN_0\delta N\frac{\rmd r}{r}=0.
\end{equation}
Noting that $N_0$ varies by a maximum of $\mathcal{O}\left(\Delta N\right)$ about $N_0^\star$ within the interval where $\delta N$ differs from zero, we find that the smooth component in equation~(\ref{eq:limit_step}) scales with $(\delta N/N)^2$, as per our choice of the reference model.   

With the reference model defined as above one then finds
\begin{equation}
    \frac{\delta N}{N_0}\approx-\frac{\Delta N^\star}{N_0^\star}\left[\mathcal{H}\left(\tilde\omega_{\rm g}^{r}-\tilde\omega_{\rm g}^{r^\star}\right)+\frac{\tilde\omega_{\rm g}^{a}-\tilde\omega_{\rm g}^{r}}{\Delta\omega_{\rm g}}\right],
    \label{deltaNovN}
\end{equation}
in the region where $\delta N $ is non-zero, where $\mathcal{H}$ is the Heaviside function. 

Finally, with $\delta N$ in hand, we can calculate the integral on the numerator of equation~(\ref{eq:periodpert_ap}) and the respective $\delta P$. Specifically,
\begin{equation}
    \delta I=\frac{\omega}{2\pi^2}\tilde\Psi_0^2\int_{\tilde\omega_{\rm g,0}^{\rm a}}^{\tilde\omega_{\rm g,0}^{\rm b}}\frac{\delta N}{N_0}\sin^2\left(\frac{\tilde\omega_{\rm g,0}^r}{\omega}+\frac{\pi}{4}+\tilde\delta\right)\rmd\tilde\omega_{\rm g,0}^r.
    \label{dI}
\end{equation}
Integrating once by parts and noting that $\rmd/\rmd \omega_{\rm g}^r=\rmd/\rmd \omega_{\rm g,0}^r(1+\mathcal{O}(\delta N/N_0))$ and $\Delta\omega_{\rm g}=\Delta\omega_{\rm g,0}(1+\mathcal{O}(\delta N/N_0))$, we find
{\begin{equation}
    \delta I\approx-\frac{\omega^2}{8\pi^2}\tilde\Psi_0^2\left[\frac{\Delta N}{N_0}\right]_{r^\star}\left[1-\frac{\omega}{\Delta\omega_{\rm g}}\sin\left(\frac{\Delta\omega_{\rm g}}{\omega}\right)\right]\sin\left(2\tilde\beta_2\right).
    \label{dI2}
\end{equation}}
Recalling that $\Delta\omega_{\rm g}\gg\omega$ and that $\omega_{\rm g}=2\pi^2/\Delta P_{\rm{as}}$,  we then find, to first order in $\delta N/N_0$, that
{\begin{equation}
    \delta P\approx \frac{\Delta P_{\rm as}}{2\pi}\left[\frac{\Delta N}{N_0}\right]_{r^\star}\sin\left(2\tilde\beta_2\right),
    \label{eq:step_lim_vp}
\end{equation}}
which is equivalent to equation~(\ref{eq:limit_step}) for the case of a zero smooth contribution, as expected.

\subsection{Small Gaussian-like glitch}
\label{apa_G}
 In the case of a Gaussian-like glitch, the reference model is defined by the buoyancy frequency in the absence of the Gaussian perturbation. Thus, the perturbation relative to that reference is simply
\begin{equation}
	\frac{\delta N}{N_0}=\frac{\Delta N}{N_0}\approx\frac{A_{\rm G}}{\sqrt{2\pi}\Delta_{\rm g}}\exp{\left(-\frac{(\omega_{\rm g}^r-\omega_{\rm g}^{\star})^2}{2\Delta_{\rm g}^2}\right)}.
	\end{equation}
Substituting in equation~(\ref{dI}) and combining with equation~(\ref{I0}), one finds, from equation(~\ref{eq:periodpert_ap}),
{\begin{equation}
    \delta P\approx -P_0\frac{\Delta P_{\rm as}}{2\pi^2}A_G+\frac{\Delta P_{\rm as}}{\pi}A_G{\it f}_\omega^{\Delta_g}\cos\left(2\beta_2\right),
    \label{eq:gaus_lim_vp}
\end{equation}
where we neglected the difference between $\omega_{\rm g,0}$ and $\omega_{\rm g}$ in the definition of $\beta_2$, as this difference is $\sim\mathcal{O}(\delta N/N_0)$.}

\section{Marginalised distributions}
\label{apB}
Here we provide the priors (Table~\ref{tab:prior}) and marginalised distributions for the fits of equation~(\ref{period_as}) to the model data, considering the step-like glitch (Fig.~\ref{fig:corner_step}) and Gaussian-like glitch  (Fig.~\ref{fig:corner_gaussian}). 

\FloatBarrier
\begin{table}
	\caption{{Lower and upper limits of the uniform prior distributions applied in the fit of equation~(\ref{period_as}) to model data, for the cases of a step-like glitch (equation~(\ref{phi_minus})) and Gaussian-like glitch (equation~(\ref{phi_gau})).}}
	\label{tab:prior}
	\begin{minipage}{0.48\textwidth}
 \centering
		\resizebox{0.7\linewidth}{!}{%
			\begin{tabular}{|c|c|c|}
   \hline				
Parameter	&	Step-like	&	Gaussian-like \\
\hline									
\multirow{2}{*}{$P_{\rm s,min}$ (s)}	&	49000	&	14900		\\
& 56000 & 15000\\
\hline									
\multirow{2}{*}{$\Delta P_{\rm as}$ (s)} &	7000	&	65 \\
& 10000 & 69\\
\hline									
\multirow{2}{*}{$A_{\rm st}$ or $A_{\rm G}$ (10$^{-6}$ rad/s)}	&	1	&	300\\
& 15 & 800\\
\hline									
\multirow{2}{*}{$\omega_{\rm g}^\star$ (10$^{-6}$ rad/s) }	&	300 &	1500 \\
&400 & 2000\\
\hline									
\multirow{2}{*}{$\Delta_{\rm g}$ (10$^{-6}$ rad/s)}	&	--&	100\\
& -- & 200\\
\hline
\multirow{2}{*}{$\delta$}	& -$\pi$/2 &	-$\pi$/2 \\
& $\pi$/2 &	$\pi$/2 \\
\hline									
\multirow{2}{*}{$\rm{jiter} \equiv \sigma$ (s)} & 0 & 0 \\
& $\inf$ & $\inf$ \\
\hline							
\end{tabular}}	
	\end{minipage}
\end{table}

\begin{figure*}
\includegraphics[width=2\columnwidth]{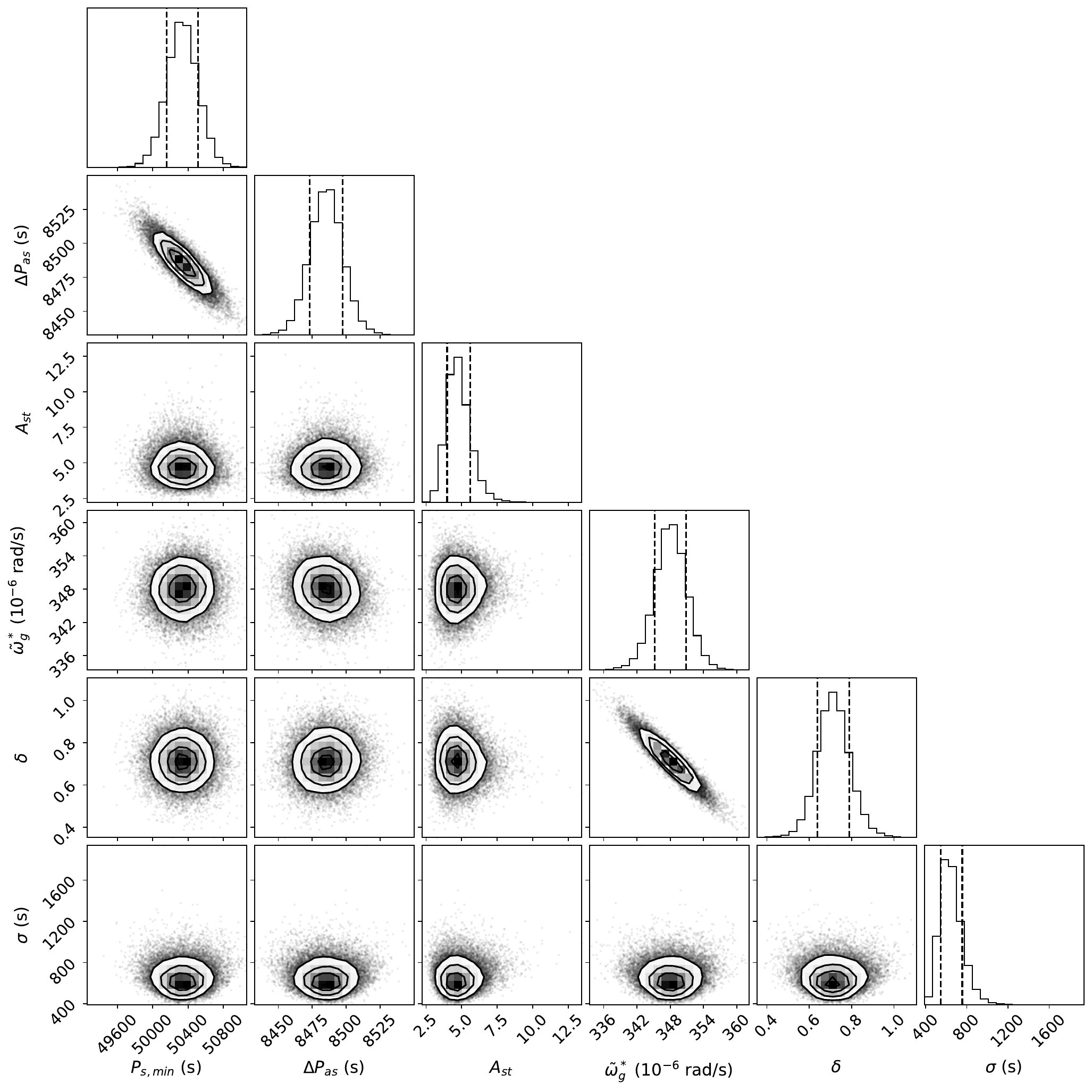}
   \caption{Marginalised distributions for the parameters considered in the fit to model data of the expression for the step-like glitch (equation~(\ref{period_as}) with $\Phi$ defined by equation~(\ref{phi_minus})). }
    \label{fig:corner_step}
\end{figure*}

\begin{figure*}
\includegraphics[width=2\columnwidth]{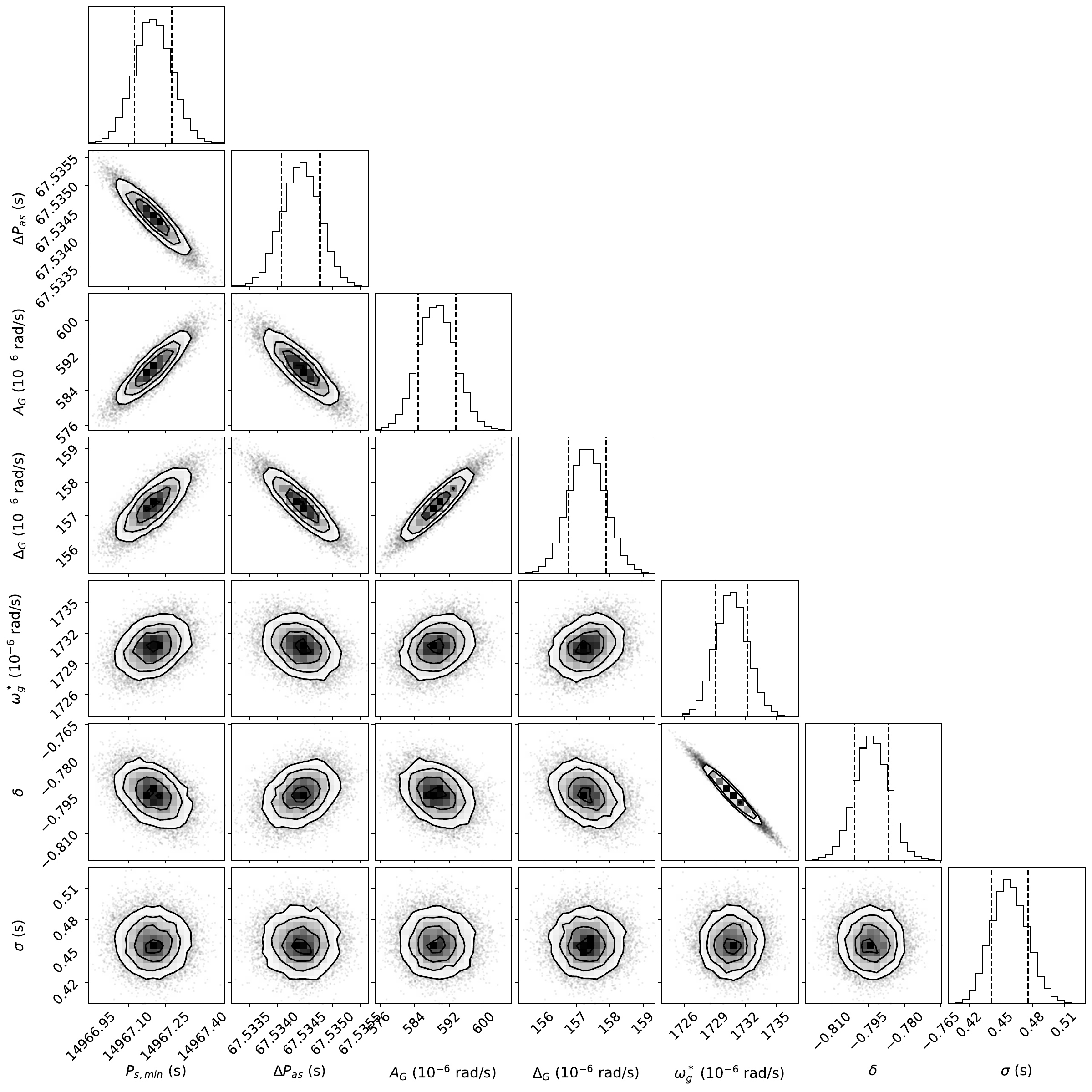}
   \caption{Marginalised distributions for the parameters considered in the fit to model data of the expression for the Gaussian-like glitch (equation~(\ref{period_as}) with $\Phi$ defined by equation~(\ref{phi_gau})). }
    \label{fig:corner_gaussian}
\end{figure*}

\section{Glitch phase $\Phi$ for mixed modes}
\label{apC}
Here we discuss how the glitch phase $\Phi$ is modified when considering mixed modes instead of pure gravity modes. To establish the eigenvalue condition in the presence of a glitch, \cite{cunha15} and \cite{cunha19} started from equation (\ref{waveeqap1}), considering the asymptotic solutions on either side of the glitch. When studying the impact on pure g modes (\ie in the absence of mode coupling),  well inside the g-mode cavity, the solutions inwards and outwards from the glitch location are, respectively,
\begin{equation}
\Psi_{\rm in} \sim \tilde\Psi_{\rm in} K_{\rm in}^{-1/2}
\sin\left(\int_{r_1}^r K_{\rm in}\rmd r +
  \frac{\pi}{4}\right),
\label{gasymp_l}
\end{equation}
and
\begin{equation}
\Psi_{\rm out} \sim \tilde\Psi_{\rm out} K_{\rm out}^{-1/2}
\sin\left(\int_r^{r_2} K_{\rm out}\rmd r +
  \frac{\pi}{4}\right),
\label{gasymp_r}
\end{equation}
where $\tilde\Psi_{\rm in}$ and $\tilde\Psi_{\rm out}$ are constants
and $K_{\rm in}$ and $K_{\rm out}$ refer again to $K$ computed inwards and outwards from the glitch location, respectively.

In the case of mixed modes, the solution outwards from the glitch position is modified due to mode coupling. That can be accounted for by introducing a coupling phase $\varphi$ in equation~(\ref{gasymp_r}) that accounts for the impact of the coupling. In that case  \citep{cunha19},
\begin{equation}
\Psi_{\rm out} \sim \tilde\Psi_{\rm out} K_{\rm out}^{-1/2}
\sin\left(\int_r^{r_2} K_{\rm out}\rmd r +
  \frac{\pi}{4}+\varphi\right),
\label{gasymp_r_c}
\end{equation}
where the frequency dependent coupling phase is given by 
\begin{equation}
\varphi={\rm atan}\left[\frac{q}{\tan\left[\left(\omega-\omega_{{\rm a},n}\right)/\omega_{\rm p}\right]}\right],
\label{varphi}
\end{equation}
with
\begin{equation}
\omega_{\rm p}=\left(\int_{r_3}^{r_4}c^{-1}\rmd r\right)^{-1}\approx 2\Delta\nu.
\label{omega_p}
\end{equation} 
Here, $r_3$ and $r_4$ are the turning points of the p-mode cavity and $q$ is the coupling coefficient~\citep{unno89,takata16}.
Also, $\omega_{{\rm a},n}$ is the angular frequency of what would be the pure acoustic mode of (pressure) radial order $n$, in the absence of mode coupling.
Based on the asymptotic solutions and the conditions obeyed by them across the glitch, the authors derived eigenvalue conditions appropriate for each case under study. 

To proceed, we consider the case of a decreasing step-like buoyancy glitch as an example.  It is straightforward to show that the conclusions remain the same for other shapes of  the buoyancy glitch, by performing a similar analysis.

The eigenvalue condition derived by \cite{cunha19} for a decreasing step-like buoyancy glitch dropping by $A_{\rm st}=N_{\rm in}/N_{\rm out}-1$ at $r^\star$in the absence of coupling is
\begin{eqnarray}
\sin\left(\int_{r_1}^{r_2}K\rmd 
  r+\frac{\pi}{2}\right)= && \nonumber\\
-A_{\rm st}
\sin\left(\int_{r^\star}^{r_2}K_{\rm out}\rmd
  r+\frac{\pi}{4}\right)\cos\left(\int_{r_1}^{r^\star}K_{\rm in}\rmd
  r+\frac{\pi}{4}\right).
\label{eigen_step_ap} 
\end{eqnarray}

Following the same analysis, but replacing equation~(\ref{gasymp_r}) by equation~(\ref{gasymp_r_c}), it follows that the eigenvalue condition becomes
\begin{eqnarray}
\sin\left(\int_{r_1}^{r_2}K\rmd 
  r+\frac{\pi}{2}+\varphi\right)= && \nonumber\\
-A_{\rm st}
\sin\left(\int_{r^\star}^{r_2}K_{\rm out}\rmd
  r+\frac{\pi}{4}+\varphi\right)\cos\left(\int_{r_1}^{r^\star}K_{\rm in}\rmd
  r+\frac{\pi}{4}\right).
\label{eigen_step_c}
\end{eqnarray}

We now consider separately the cases of a glitch in the inner half and the outer half of the g-mode cavity.
For a glitch in the inner half of the g-mode cavity one has
\begin{equation}
\int_{r^\star}^{r_2} K_{\rm out}\rmd r+\frac{\pi}{4} +\varphi = \int_{r_1}^{r_2} K\rmd r+\frac{\pi}{2}+\varphi-\int_{r_1}^{r^\star} K_{\rm in}\rmd r-\frac{\pi}{4},
\end{equation}
and substituting in Eq.~(\ref{eigen_step_c}) one finds
\begin{eqnarray}
\sin\left(\int_{r_1}^{r_2}K\rmd  r+\frac{\pi}{2}+\Phi+\varphi\right) = 0,
\label{eigengap2}
\end{eqnarray}
where $\Phi$, and a new quantity, $B$, are defined by the following system of equations,
\begin{equation}
\left\{
\begin{array}{lll}
B\cos\Phi \hspace{-0.0cm}&\hspace{-0.0cm}= & \hspace{-0.0cm} 1+A_{\rm st}\cos^2\left(\int_{r_1}^{r^\star} K_{\rm in}\rmd r+\frac{\pi}{4}\right),
\\ \\
B\sin\Phi\hspace{-0.0cm}&\hspace{-0.0cm}= &\hspace{-0.0cm}
-\frac{1}{2}A_{~\rm st}\cos\left(2\int_{r_1}^{r^\star} K_{\rm in}\rmd r\right).
\end{array}
\right.
\label{Phi_st_in}
\end{equation}
By comparing with equation~(\ref{phi_minus}), we thus conclude that in the case of a glitch located in the inner half of the g-mode cavity, the glitch phase $\Phi$ entering the perturbation to the mixed-mode frequencies is the same as that derived for the pure g modes.

For a glitch in the outer half of the g-mode cavity one has instead
\begin{equation}
\int_{r_1}^{r^\star} K_{\rm in}\rmd r+\frac{\pi}{4} = \int_{r_1}^{r_2} K\rmd r+\frac{\pi}{2}+\varphi-\int_{r^\star}^{r_2} K_{\rm out}\rmd r-\frac{\pi}{4}-\varphi.
\end{equation}
Substituting in equation~(\ref{eigen_step_c}), we find again equation~(\ref{eigengap2}), but with the glitch phase now defined by the system of equations
\begin{equation}
\left\{
\begin{array}{lll}
B\cos\Phi \hspace{-0.0cm}&\hspace{-0.0cm}= & \hspace{-0.0cm} 1+A_{\rm st}\sin^2\left(\int_{r^\star}^{r_2} K_{\rm out}\rmd r+\frac{\pi}{4}+\varphi\right),
\\ \\
B\sin\Phi\hspace{-0.0cm}&\hspace{-0.0cm}= &\hspace{-0.0cm}
\frac{1}{2}A_{~\rm st}\cos\left(2\int_{r^\star}^{r_2} K_{\rm out}\rmd r+2\varphi\right).
\end{array}
\right.
\label{Phi_st_ou}
\end{equation}
In this case the glitch phase can be written as,
\begin{eqnarray}
    \Phi={\rm arccot}\left[\frac{2}{A_{\rm st}\cos{\left(2\int_{r^\star}^{r_2} K_{\rm out}\rmd r+2\varphi\right)}}\right. \nonumber\\+\left.\tan\left(\int_{r^\star}^{r_2} K_{\rm out}\rmd r+\frac{\pi}{4}+\varphi\right)\right].
        \label{phi_st_coupling}
\end{eqnarray}
To see how the glitch phase is modified for the mixed modes in this case, one can compare equation~(\ref{phi_st_coupling}) to equation~(\ref{phi_minus}).
Recalling that $\Delta N=N_{\rm in}-N_{\rm ou}$, we rewrite $A_{\rm st}=N_{\rm in}/N_{\rm ou}-1=\Delta N/(N_{\rm in}-\Delta N)$. Using $A_{\rm st}$ expressed in terms of $\Delta N$ in equation~(\ref{phi_st_coupling}) and noting that $\tan(\theta)+\cot(\theta)= 1/(\sin(\theta)\cos(\theta))$ for any $\theta$, equation~(\ref{phi_st_coupling}) can be rewritten as
\begin{eqnarray}
    \Phi={\rm arccot}\left[-\frac{2}{\hat A_{\rm st}\cos{\left(2\int_{r^\star}^{r_2} K_{\rm out}\rmd r+2\varphi\right)}}\right.\nonumber\\ -\left.\cot\left(\int_{r^\star}^{r_2} K_{\rm out}\rmd r+\frac{\pi}{4}+\varphi\right)\right].
        \label{phi_st_coupling2}
\end{eqnarray}
Equation~(\ref{phi_st_coupling2}) has the same functional form as equation~(\ref{phi_minus}), and we note that the appearance of $\hat A_{\rm st}$ instead of $A_{\rm st}$ was anticipated given that by symmetry a decreasing step-like glitch in the outer cavity should imprint the same signature on pure g modes as an increasing step-like glitch in the inner g-mode cavity, with an associated positive amplitude defined by $\hat A_{\rm st}$. Nevertheless, there is a second important difference that we can identify when comparing the two results. In the case of the mixed modes and a glitch located in the outer half of the g-mode cavity, the arguments of the sinusoidal functions change, incorporating, in addition, the coupling phase, $\varphi$. This means that the coupling between the two cavities will impact the glitch phase when the glitch is located in the outer half of the cavity.  To recover the same expression for $\Phi$ as in the pure g-mode case, one must now include the coupling phase $\varphi$ in the definition of $\beta_2$.


\end{appendix}
\end{document}